\documentclass[11pt]{article}
\usepackage[percent]{overpic}
\usepackage{kotex}
\usepackage{fullpage} 
\usepackage{setspace} 
\usepackage[utf8]{inputenc}
\usepackage{amsmath}
\usepackage{amsfonts}
\usepackage{amssymb}
\usepackage{amsthm}
\usepackage{authblk}
\usepackage{mathtools}
\usepackage{float}
\usepackage{accents}
\usepackage{booktabs}
\usepackage[round]{natbib}    
\usepackage{xr}                
\usepackage{xcolor}        
\usepackage{placeins}
\usepackage{enumitem}        
\usepackage{algorithm}       
\usepackage{algpseudocode}  
\usepackage[colorlinks=true, linkcolor=blue, urlcolor=blue, citecolor=blue]{hyperref}

\newtheorem{theorem}{Theorem}

\newcommand{\indep}{\perp \!\!\! \perp}

\title{$\ell_0$-Regularized Item Response Theory Model\\ for Robust Ideal Point Estimation}

\author[1]{Kwangok Seo}
\author[1]{Johan Lim}
\author[2]{Seokho Lee}
\author[3]{Jong Hee Park}

\affil[1]{Department of Statistics, Seoul National University, Seoul, Korea}
\affil[2]{Department of Statistics, Hankuk University of Foreign Studies, Yongin, Korea}
\affil[3]{Department of Political Science and International Relations, Seoul National University, Seoul, Korea}

\date{}

\begin{document}

\maketitle 

\begin{abstract}
    Ideal point estimation methods face a significant challenge when legislators engage in protest voting -- strategically voting against their party to express dissatisfaction. Such votes introduce attenuation bias, making ideologically extreme legislators appear artificially moderate. We propose a novel statistical framework that extends the fast EM-based estimation approach of \cite{Imai2016} using $\ell_0$ regularization method to handle protest votes. Through simulation studies, we demonstrate that our proposed method maintains estimation accuracy even with high proportions of protest votes, while being substantially faster than MCMC-based methods. Applying our method to the 116th and 117th U.S. House of Representatives, we successfully recover the extreme liberal positions of ``the Squad'', whose protest votes had caused conventional methods to misclassify them as moderates. While conventional methods rank Ocasio-Cortez as more conservative than 69\% of Democrats, our method places her firmly in the progressive wing, aligning with her documented policy positions. This approach provides both robust ideal point estimates and systematic identification of protest votes, facilitating deeper analysis of strategic voting behavior in legislatures.

\end{abstract}

\baselineskip 18pt 

\section{Introduction}
Ideal point estimation has become an essential statistical tool for understanding legislative behavior and political preferences \citep{Poole1997, Bailey2001, Martin2002, Clinton2004, Spirling2010, Peress2010, Royce2013, Imai2016, Goplerud2019, Moser2021, Binding2022}. While methodological advances have made these estimates increasingly sophisticated, a fundamental challenge remains: how to handle protest votes - strategic anti-partisan votes cast by legislators against their own party to express dissatisfaction with party leadership or specific policies. This paradox becomes particularly acute when studying ideologically extreme legislators, who despite typically being the strongest party-line voters, occasionally break ranks through protest votes that can distort estimates of their true ideological positions. Such strategic voting behavior challenges core assumptions of spatial voting models and raises questions about how to accurately measure legislator ideology when votes reflect both sincere policy preferences and tactical political calculations.

Consider ``the Squad" in the U.S. House of Representatives - a group of progressive Democratic legislators including Alexandria Ocasio-Cortez (NY), Ilhan Omar (MN), Ayanna Pressley (MA), and Rashida Tlaib (MI). These members champion policies like the Green New Deal and advocate abolishing ICE, positions that place them firmly on the progressive wing of the Democratic Party. Yet curiously, conventional ideal point estimation methods like NOMINATE and Bayesian item response theory (BIRT) models characterize Squad members as relative moderates. NOMINATE's ranking of Ocasio-Cortez, which is available in Voteview \citep{voteview2022}, as ``more conservative than 69\% of Democrats in the 116th House'' stands in stark contrast to her well-documented progressive policy positions.

This apparent contradiction stems from the Squad's strategic use of protest votes, which have become increasingly frequent and consequential in recent Congresses. In November 2021, all six Squad members (now including Cori Bush and Jamaal Bowman) joined conservative Republicans in opposing the bipartisan infrastructure bill—not from ideological alignment, but to protest what they viewed as insufficient commitment to broader progressive priorities in the Build Back Better Act. This high-stakes protest vote was designed to preserve leverage for progressive priorities like childcare, healthcare, and climate action, even as it placed them in statistical alignment with conservative Republicans in roll-call analyses.

Even more striking was the May 2021 Capitol Security funding bill, which nearly failed due to progressive protest votes. With a final tally of 213-212, three Squad members (Bush, Omar, and Pressley) voted ``No'' while three others (Bowman, Ocasio-Cortez, and Tlaib) voted ``Present'' to register their objection to increased police funding without addressing ``underlying threats.'' With Republicans unanimously opposed, these strategic protest votes pushed the bill to the very edge of failure, demonstrating how consequential such voting behavior has become in narrowly divided Congresses. While political observers readily recognize votes like Ocasio-Cortez's opposition to the Capitol Security funding bill as protests rather than indicators of ideological moderation, current statistical methods struggle to properly account for this increasingly common strategic behavior.

Prior research has made significant progresses in addressing non-sincere voting behavior through various methodological innovations. Notable contributions include IRT models with fatter tails \citep{Gelman2005} and flexible item characteristic curves \citep{Montgomery2020, Montgomery2023}. Building on these advances, we propose a general ideal point estimation method that not only preserves the capabilities of conventional approaches but also introduces the ability to identify and account for protest votes. Specifically, our method introduces a shift parameter to identify ``non-conforming" votes within a legislator's overall voting pattern. By employing $\ell_0$ regularization across all votes, we systematically detect protest votes while maintaining model parsimony. 

The theoretical foundation of our approach draws from robust statistics, particularly the concept of shift parameters for identifying atypical patterns in data. This methodology was pioneered by \cite{mccann2007} in the context of robust regression. A significant advancement came from \cite{she2011}, who demonstrated that while $\ell_1$ regularization of shift parameters yields Huber's M-estimate, $\ell_0$ regularization proves more effective at accurate outlier identification and model robustification. The shift parameter framework has been successfully extended to various domains. \cite{lee2012} applied it to margin-based loss minimization, including logistic regression and support vector machines, while \cite{lee2016} explored its application with functional covariates. \cite{shin2023} provided theoretical justification for preferring $\ell_0$ over $\ell_1$ regularization in robust logistic regression. The versatility of this approach is further evidenced by \cite{witten2013}'s extension to unsupervised learning contexts, including principal component analysis and $k$-means clustering. 

Regularization approaches have become an increasingly important methodological direction in IRT modeling for making estimation feasible and enhancing interpretation. Recent work, such as \cite{robitzsch2024smooth}, provides an overview of these developments and illustrates their application in various IRT frameworks.

For efficient estimation, we adopt \cite{Imai2016}'s EM-based strategy, enabling reliable computation with large-scale legislative data.

The remainder of the paper is organized as follows: In Section \ref{sec_2}, we briefly review the BIRT model proposed by \cite{Clinton2004}, focusing on the one-dimensional case, which serves as the basis for our proposed model. In Section \ref{sec_3}, we extend the one-dimensional BIRT model by introducing shift parameters with priors specified in \eqref{eqn:prior_gamma}  (Section \ref{sub_sec_3.1}), and we develop an expectation-maximization (EM) algorithm to estimate the proposed model (Section \ref{sub_sec_3.2}). In Section \ref{sec_4}, we justify the prior assumption for the introduced shift parameters by illustrating its connection to the spike-and-slab prior (Theorem \ref{theorem1}, Section \ref{sub_sec_4.1}). We then discuss the identifiability issue of our proposed model (Theorem \ref{theorem2}, Section \ref{sub_sec_4.2}). In Sections \ref{sec_5} and \ref{sec_6}, we conduct simulation studies and analyze roll-call data from the 116th and 117th U.S. House of Representatives, respectively, to evaluate the performance of the proposed method. In Section \ref{sec_7}, we extend our proposed method to the multidimensional case, and conclude the paper by outlining directions for future research in Section \ref{sec_8}.

\section{Preliminary: BIRT Model} \label{sec_2}
Consider a set of $ I $ legislators voting on $ J $ bills in a legislative body. For each legislator $ i $ and bill $ j $, we observe a binary vote $ y_{ij} \in \{0,1\} $, where 1 indicates a yea vote and 0 indicates a nay vote. The canonical spatial voting model in political science \cite[e.g.][]{Poole1997, Clinton2004, Royce2013} explains legislators' decisions using a utility maximization framework.
Specifically, the utilities of voting yea ($ U_{ij}^{\text{yea}} $) and nay ($ U_{ij}^{\text{nay}} $) are defined as follows:
\begin{align}\label{eqn:utility_yea_nay}
\begin{split}
    U_{ij}^{\text{yea}} &= U^{\text{yea}}(\theta_i, \zeta_j) = -(\theta_i - \zeta_j)^2 + \eta_{ij},\\
    U_{ij}^{\text{nay}} &= U^{\text{nay}}(\theta_i, \psi_j) = -(\theta_i - \psi_j)^2 + \nu_{ij},
\end{split}
\end{align}  
where $ \theta_i \in \mathbb{R} $ represents legislator $ i $'s ideal point in the policy space, while $ \zeta_j \in \mathbb{R}$ and $ \psi_j \in \mathbb{R}$ represent the policy positions associated with yea and nay votes for bill $ j $, respectively. The terms $ \eta_{ij} $ and $ \nu_{ij} $ are stochastic errors, assumed to satisfy $ \eta_{ij} - \nu_{ij} \sim N(0, \sigma_j^2) $, where $ N(\mu, \sigma^2) $ denotes a normal distribution with mean $ \mu $ and variance $ \sigma^2 $.   
The utility of legislator $i$ for bill $j$ is then given by  
\begin{align}\label{eqn:utility_ij}
    U_{ij} = U(\theta_i, \zeta_j, \psi_j)  = U_{ij}^{\text{yea}} - U_{ij}^{\text{nay}} = -(\theta_i - \zeta_j)^2 + (\theta_i - \psi_j)^2 + (\eta_{ij} - \nu_{ij}).
\end{align}  
By the normality assumption for stochastic error terms, we have
\begin{align*}
    p(U_{ij} \geq 0 \mid \theta_i, \zeta_j, \psi_j, \sigma_j) 
    &= p\left(Z_{ij} \geq \frac{(\theta_i - \zeta_j)^2 - (\theta_i - \psi_j)^2}{\sigma_j} ~\Bigg|~ \theta_i, \zeta_j, \psi_j, \sigma_j\right)\\
    &= p\left(Z_{ij} \leq  \frac{\psi_j^2 - \zeta_j^2}{\sigma_j} + \frac{2(\zeta_j - \psi_j)}{\sigma_j}  \theta_i ~\Bigg |~ \theta_i, \zeta_j, \psi_j, \sigma_j\right)\\
    &= p(Z_{ij} \leq \alpha_j + \beta_j  \theta_i \mid \alpha_j, \beta_j, \theta_i)\\
    &= \Phi(\alpha_j + \beta_j \theta_i),
\end{align*}  
where $ Z_{ij} $ denotes a standard normal random variable, $ \alpha_j = (\psi_j^2 - \zeta_j^2)/\sigma_j $ represents the difficulty parameter, and $ \beta_j = 2(\zeta_j - \psi_j)/\sigma_j $ is the discrimination parameter. Additionally, $\Phi(\cdot)$ denotes the distribution function of the standard normal random variable. 

To connect the voting outcome $ Y_{ij} $ with the corresponding utility $ U_{ij} $, \cite{Clinton2004} proposed modeling the probability of $ Y_{ij} $ being a yea vote as
\begin{align*}
    p(Y_{ij} = 1 \mid \alpha_j, \beta_j, \theta_i) = p(U_{ij} \geq 0 \mid \theta_i, \zeta_j, \psi_j, \sigma_j).
\end{align*}
This formulation is intuitive since a positive value of $U_{ij}$ indicates that voting yea provides greater utility than voting nay, while a negative value implies the opposite.
The likelihood of the observed voting outcomes is then given by
\begin{align}\label{eqn:like}
    \prod_{i=1}^I \prod_{j=1}^J p(y_{ij} \mid \alpha_j, \beta_j, \theta_i) = \prod_{i=1}^I \prod_{j=1}^J \Phi(\alpha_j + \beta_j \theta_i)^{y_{ij}} \cdot \{1-\Phi(\alpha_j + \beta_j \theta_i)\}^{1-y_{ij}}.
\end{align}  
With the following prior distributions for the model parameters
\begin{align}\label{eqn:prior_abt}
    \boldsymbol{\tilde{\beta}} = (\alpha_j, \beta_j)^\top \sim N(\boldsymbol{\mu}_{\boldsymbol{\tilde{\beta}}}, \Sigma_{\boldsymbol{\tilde{\beta}}}), \quad \theta_i \sim N(\mu_{\theta}, \sigma_{\theta}^2),
\end{align}  
and the data augmentation scheme of \cite{AlbertChib1993}, which introduce an auxiliary latent variable $ y_{ij}^* $\footnote{To facilitate the subsequent MCMC sampling step, \cite{Clinton2004} adopt the data augmentation scheme proposed by \cite{AlbertChib1993} and introduce the latent variable $ y_{ij}^* $. Specifically, they assume that  
\begin{align*}
y_{ij}^* \mid \alpha_j, \beta_j, \theta_i \sim N(\alpha_j + \beta_j \theta_i, 1),\quad 
Y_{ij} \mid y_{ij}^*, \alpha_j, \beta_j, \theta_i \sim \delta_1(y_{ij}) \cdot I(y_{ij}^* \geq 0) + \delta_0(y_{ij}) \cdot I(y_{ij}^* < 0).
\end{align*}
By integrating out the latent variable $ y_{ij}^* $, we recover the conditional distribution of $ Y_{ij} $ given $ \alpha_j $, $ \beta_j $, and $ \theta_i $, which follows a Bernoulli distribution with $p(Y_{ij} = 1) = \Phi(\alpha_j + \beta_j \theta_i)$.}, they derive the following posterior distribution:
\begin{align}\label{eqn:post_BIRT}
    \begin{split}
        p\left(\{y_{ij}^*\}, \{\theta_i\}, \{\alpha_j\}, \{\beta_j\} \mid \{y_{ij}\}\right) 
        &\propto \prod_{i=1}^I \prod_{j = 1}^J 
        \bigg[\left(\delta_1(y_{ij})  \cdot \mathbb{I}(y_{ij}^* \geq 0) + \delta_0(y_{ij}) \cdot \mathbb{I}(y_{ij}^* < 0) \right) \\
        &\quad \quad \times \phi_1\left(y_{ij}^*; \boldsymbol{\tilde{\beta}}_j^\top \boldsymbol{\tilde{\theta}}_i, 1\right) \times \phi_{1}\left(\theta_i; \mu_{\theta}, \sigma_{\theta}^2 \right) \times \phi_{2}\left(\boldsymbol{\tilde{\beta}}_j ; \boldsymbol{\mu}_{\boldsymbol{\tilde{\beta}}}, \Sigma_{\boldsymbol{\tilde{\beta}}}\right)\bigg],
    \end{split}
\end{align}  
where $ \delta_x(\cdot) $ denotes the Dirac delta function, which places a point mass at $ x $, and $\boldsymbol{\boldsymbol{\tilde{\theta}}}_i = (1, \theta_i)^\top$. The function $ \phi_k(\cdot ; \boldsymbol{\mu}, \Sigma) $ represents the density of a $ k $-dimensional normal distribution with mean $ \boldsymbol{\mu} $ and variance  $ \Sigma $.  

Based on this posterior distribution, \cite{Clinton2004} develop an MCMC sampler to obtain posterior samples of the ideal points. To mitigate the computational burden of the fully Bayesian approach, \cite{Imai2016} propose using the EM algorithm, which maximizes the same posterior distribution to obtain the maximum a posteriori (MAP) estimates of the ideal points.

\section{Proposed Method}\label{sec_3}
In this section, we extend the BIRT model by introducing shift parameters, $\gamma_{ij}$, along with their corresponding priors, and provide the rationale and necessity for this extension (Section \ref{sub_sec_3.1}). We then propose an EM algorithm that maximizes the posterior in \eqref{eqn:post_l0}  to obtain the maximum a posteriori (MAP) estimates of our parameters of interest (Section \ref{sub_sec_3.2}).

\subsection{Robust BIRT Model and Strategic Voting 
Behavior}\label{sub_sec_3.1}
We adopt the definition of the utility for voting yea and nay as described in \eqref{eqn:utility_yea_nay}. However, we introduce an additional parameter $ \xi_{ij} $ in the utility $ U_{ij} $ to account for strategic voting behaviors. Specifically, we define the utility as  
\begin{align}\label{eqn:utility_ij_xi}
    U_{ij} = U(\theta_i, \zeta_j, \psi_j, \xi_{ij}) = U_{ij}^{\text{yea}} - U_{ij}^{\text{nay}} + \xi_{ij} = -(\theta_i - \zeta_j)^2 + (\theta_i - \psi_j)^2 + \xi_{ij} + (\eta_{ij} - \nu_{ij}).
\end{align}  
Given the definition of $U_{ij}$ in \eqref{eqn:utility_ij_xi} and the normality assumption for stochastic errors, we can express 
\begin{align*}
    p(U_{ij} \geq 0 \mid \theta_i, \zeta_j, \psi_j, \sigma_j, \xi_{ij}) 
    &= \Phi(\alpha_j + \beta_j \theta_i + \gamma_{ij}),
\end{align*}  
where $\gamma_{ij} = \xi_{ij}/\sigma_j$ represents the shift parameter.
Following the \cite{Clinton2004}, we model the probability of $Y_{ij}$ being a yea votes as 
\begin{align*}
     p(Y_{ij} = 1 \mid \alpha_j, \beta_j, \theta_i, \gamma_{ij}) = p(U_{ij} \geq 0 \mid \theta_i, \zeta_j, \psi_j, \sigma_j, \xi_{ij}),
\end{align*}
which implies that the likelihood of the observed voting outcomes is
\begin{align}\label{eqn:like_gamma}
    \prod_{i=1}^I \prod_{j=1}^J p(y_{ij} = 1 \mid \alpha_j, \beta_j, \theta_i, \gamma_{ij}) = \prod_{i=1}^I \prod_{j = 1}^J \Phi(\alpha_j + \beta_j \theta_i + \gamma_{ij})^{y_{ij}} \cdot \{1-\Phi(\alpha_j + \beta_j \theta_i + \gamma_{ij})\}^{1-y_{ij}}.
\end{align}  
It is notable that if all $ \gamma_{ij} $ are equal to 0 in \eqref{eqn:like_gamma}, our model reduces to the model \eqref{eqn:like} considered in \cite{Clinton2004}.
The role of the shift parameter $ \gamma_{ij} $ is to identify strategic votes and mitigate their impact on the estimation of bill parameters $ \alpha_j, \beta_j $ and ideal points $ \theta_i $. Specifically, for the sincere votes (the non-strategic votes), $\gamma_{ij} $ should be set to 0, allowing the observed voting outcomes to be directly reflected in estimating ideal points. Conversely, for the strategic votes, $ \gamma_{ij} $ should take a non-zero value to bridge the gap between the actual voting outcome and the legislator's ground truth utility for that vote, thereby ensuring that strategic votes do not distort the estimation of bill parameters and ideal points.

We adopt assumption of prior for $\boldsymbol{\tilde{\beta}}_j$ and $\theta_i$ as in \eqref{eqn:prior_abt}. To complete the fully Bayesian approach, the specification of a prior for $ \gamma_{ij} $ is necessary. Specifically, we consider 
\begin{align}\label{eqn:prior_gamma}
    p(\gamma_{ij} \mid \lambda) \propto \exp\left( -\frac{\lambda^2}{2} \mathbb{I}(\gamma_{ij} \neq 0)\right),
\end{align} 
where $ \lambda $ is a hyperparameter that controls the sparsity of $ \Gamma $. Here, $ \Gamma $ is an $ I \times J $ matrix whose $(i,j)$-th element is $ \gamma_{ij} $. At first glance, the prior in \eqref{eqn:prior_gamma} may appear unconventional. We provide justification for its use in Section \ref{sub_sec_4.1}.

Putting the likelihood described in \eqref{eqn:like_gamma} and priors in \eqref{eqn:prior_abt} and \eqref{eqn:prior_gamma} together, our posterior distribution is
\begin{align}\label{eqn:post_l0}
    \begin{split}
        &p\left(\{y_{ij}^*\}, \{\theta_i\}, \{\alpha_j\}, \{\beta_j\}, \{\gamma_{ij}\} \mid \{y_{ij}\}\right) \\
        &\propto \prod_{i=1}^I \prod_{j = 1}^J 
        \bigg[\left(\delta_1(y_{ij})  \cdot \mathbb{I}(y_{ij}^* \geq 0) + \delta_0(y_{ij}) \cdot \mathbb{I}(y_{ij}^* < 0) \right) \times \phi_1\left(y_{ij}^*; \boldsymbol{\tilde{\beta}}_j^\top \boldsymbol{\tilde{\theta}}_i + \gamma_{ij}, 1\right)\\
        &\quad \quad \times \phi_1\left(\theta_i; \mu_{\theta}, \sigma_{\theta}^2 \right) \times \phi_{2}\left(\boldsymbol{\tilde{\beta}}_j ; \boldsymbol{\mu}_{\boldsymbol{\tilde{\beta}}}, \Sigma_{\boldsymbol{\tilde{\beta}}}\right) \times \exp\left( -\frac{\lambda^2}{2} \mathbb{I}(\gamma_{ij} \neq 0)\right)\bigg].
    \end{split}
\end{align}
Our goal is to find a solution that maximizes the posterior distribution described in \eqref{eqn:post_l0}.

\subsection{EM Algorithm for the Robust BIRT Model}\label{sub_sec_3.2}
To obtain a maximum a posteriori (MAP) estimates, we maximize the posterior in \eqref{eqn:post_l0} by employing the expectation-maximization (EM) algorithm. The EM algorithm iteratively updates the parameter estimates by alternating between the expectation step (E-step) and the maximization step (M-step) until convergence (or until a pre-specified convergence criterion is met). 
From now on, we assume that the parameters of interest—$ \alpha_j $, $ \beta_j $, $ \theta_i $, and $ \gamma_{ij} $ for $ i = 1, 2, \ldots, I $ and $ j = 1, 2, \ldots, J $—are fixed unknown constants, while the latent variables $ y_{ij}^* $ are treated as random variables.

Let $ \theta^{(t-1)}, \alpha_j^{(t-1)}, \beta_j^{(t-1)}, \gamma_{ij}^{(t-1)} $ denote the parameter values obtained at iteration $ t-1 $. The $ t $-th iteration starts with the E-step, where we compute the $Q$-function defined as
\begin{align}\label{eqn:q_func}
    \begin{split}
        Q(\boldsymbol{\vartheta} \parallel \boldsymbol{\vartheta}^{(t-1)}) &= \mathbb{E}\left[ \log p(\{y_{ij}^*\}, \{\theta_i\}, \{\alpha_j\}, \{\beta_j\}, \{\gamma_{ij}\} \mid \{y_{ij}\}) \mid \{y_{ij}\}, \boldsymbol{\vartheta}^{(t-1)}\right]\\
        &= -\frac{1}{2} \sum_{i = 1}^I \sum_{j = 1}^J \left( \boldsymbol{\tilde{\beta}}_j^\top \boldsymbol{\tilde{\theta}}_i \boldsymbol{\tilde{\theta}}_i^\top \boldsymbol{\tilde{\beta}}_j -2 \boldsymbol{\tilde{\beta}}_j^\top \boldsymbol{\tilde{\theta}}_i (y_{ij}^{*(t)} - \gamma_{ij})  + \gamma_{ij}^2 -2 y_{ij}^{*(t)} \gamma_{ij}\right)\\
        &\quad -\frac{1}{2} \sum_{i = 1}^I \left(\theta_i^2 /\sigma_{{\theta}}^2 - 2\theta_i \mu_{\theta}/\sigma_{\theta}^2\right) 
        -\frac{1}{2} \sum_{j = 1}^J \left(\boldsymbol{\tilde{\beta}}_j^\top \Sigma_{{\boldsymbol{\tilde{\beta}}}}^{-1} \boldsymbol{\tilde{\beta}}_j - 2\boldsymbol{\tilde{\beta}}_j^\top \Sigma^{-1}_{\boldsymbol{\tilde{\beta}}}\boldsymbol{\mu}_{\boldsymbol{\tilde{\beta}}}\right) \\
        &\quad -\frac{\lambda^2}{2} \sum_{i = 1}^I \sum_{j = 1}^J \mathbb{I}(\gamma_{ij} \neq 0) + \text{const.},
    \end{split}
\end{align}
where $\boldsymbol{\vartheta} = \{\theta_i\}\bigcup \{\alpha_j\} \bigcup \{\beta_j\}\bigcup \{\gamma_{ij}\}$, and $ y_{ij}^{*(t)} $ denotes the conditional expectation of the latent variable $y_{ij}^*$ given the parameter values at iteration $ t-1 $, $\boldsymbol{\vartheta}^{(t-1)}$. The specific form of $y_{ij}^{*(t)}$ is defined as follows:
\begin{align}\label{eqn:y_star}
\begin{split}
    y_{ij}^{*(t)} &\equiv \mathbb{E} \left(y_{ij}^* \mid \theta_i^{(t-1)}, \alpha_j^{(t-1)},  \beta_j^{(t-1)}, \gamma_{ij}^{(t-1)}, y_{ij}\right) \\
    &= \begin{cases}
        m_{ij}^{(t-1)} - \frac{\phi_1(m_{ij}^{(t-1)})}{1-\Phi(m_{ij}^{(t-1)})} & \text{if } y_{ij} = 0\\
        m_{ij}^{(t-1)} + \frac{\phi_1(m_{ij}^{(t-1)})}{\Phi(m_{ij}^{(t-1)})} & \text{if } y_{ij} = 1\\
        m_{ij}^{(t-1)} & \text{if} ~y_{ij }~ \text{is missing},
    \end{cases}
\end{split}
\end{align}
where $m_{ij}^{(t-1)} = \boldsymbol{\tilde{\theta}}_i^{(t-1)\top} \boldsymbol{\tilde{\beta}}_j^{(t-1)} + \gamma_{ij}^{(t-1)}$.

The M-step involves maximizing the $ Q $-function obtained in the E-step. It is notable that we maximize the $ Q $-function in an alternating fashion, rather than jointly maximizing it for all parameters of interest. The details are as follows: First, we maximize the $ Q $-function in \eqref{eqn:q_func} with respect to $ \{\theta_i\} $ to obtain $ \{\theta_i^{(t)}\} $. Then, we update the $ Q $-function by replacing $ \{\theta_i^{(t-1)}\} $ with $ \{\theta_i^{(t)}\} $ and maximize the updated $ Q $-function with respect to $ \{\boldsymbol{\tilde{\beta}}_j\} $, yielding $ \{\boldsymbol{\tilde{\beta}}_j^{(t)}\} $. Next, we update the $ Q $-function again by replacing $ \{\boldsymbol{\tilde{\beta}}_j^{(t-1)}\} $ with $ \{\boldsymbol{\tilde{\beta}}_j^{(t)}\} $ and maximize this updated $ Q $-function with respect to $ \{\gamma_{ij}\} $ to obtain $ \{\gamma_{ij}^{(t)}\} $. This iterative procedure ensures that each parameter is updated sequentially based on the current estimates of the others. The closed-form sequential update rule\footnote{The order of the sequential updates during the M-step is not important.} is given by
\begin{align}\label{eqn:update_rule}
\begin{split}
    \theta_i^{(t)} 
    &= \left(1/\sigma_{\theta}^{2} + \sum_{j = 1}^J \beta_j^{(t-1)2} \right)^{-1}  \left( \mu_{\theta}/\sigma_{\theta}^{2} + \sum_{j = 1}^J \beta_j^{(t-1)} \left(y_{ij}^{*(t)} - \gamma_{ij}^{(t-1)} - \alpha_j^{(t-1)}\right)\right), \\ 
    \boldsymbol{\tilde{\beta}}_j^{(t)} 
    &= \left(\Sigma_{\boldsymbol{\tilde{\beta}}}^{-1} + \sum_{i = 1}^I \boldsymbol{\tilde{\theta}}_i^{(t)} \boldsymbol{\tilde{\theta}}_i^{(t)\top} \right)^{-1}  \left( \Sigma_{\boldsymbol{\tilde{\beta}}}^{-1} \boldsymbol{\mu}_{\boldsymbol{\tilde{\beta}}} + \sum_{i = 1}^I \boldsymbol{\tilde{\theta}}_i^{(t)} \left(y_{ij}^{*(t)} - \gamma_{ij}^{(t-1)}\right)\right), \\ 
    \gamma_{ij}^{(t)} 
    &= \left( y_{ij}^{*(t)} - \boldsymbol{\tilde{\beta}}_j^{(t)\top} \boldsymbol{\tilde{\theta}}_i^{(t)}\right) ~ \mathbb{I}\left(~\Big|~y_{ij}^{*(t)} - \boldsymbol{\tilde{\beta}}_j^{(t)\top}\boldsymbol{\tilde{\theta}}_i^{(t)}~\Big|~ > \lambda\right).
\end{split}
\end{align}

The E-step and M-step alternate until convergence. We consider the algorithm to have converged when the maximum change in the shift parameters $ \gamma_{ij} $ falls below a specified tolerance $ \epsilon $, i.e.,
\begin{align}\label{con_cri}
\underset{i, j}{\max} ~\Big|~\gamma_{ij}^{(t)} - \gamma_{ij}^{(t-1)}~\Big| < \epsilon,
\end{align}
or when the maximum number of iterations is reached. 
Our EM algorithm can be seen as an extension of the framework in \cite{Imai2016}, incorporating the shift parameter $ \gamma_{ij} $ and its corresponding prior distribution, while maintaining the same basic structure.   
Algorithm \ref{alg:EM} summarizes the EM algorithm explained above. Implementation details are provided in Appendix \ref{Appen_B}.

\begin{algorithm}
    \caption{Expectation-Maximization (EM) Algorithm for the robust BIRT model}\label{alg:EM}
    \textbf{Input:}  
    \begin{itemize}
        \item $I \times J$ roll call matrix $D$
        \item Hyperparameters $\lambda$, $\mu_{\theta}$, $\sigma_\theta^2$, $\boldsymbol{\mu}_{\boldsymbol{\tilde{\beta}}}$ and $\Sigma_{\boldsymbol{\tilde{\beta}}}$.
        \item Initial estimates:  
        $\boldsymbol{\vartheta}^{(0)} = \left\{\theta_i^{(0)}\right\} \bigcup \left\{\alpha_j^{(0)}\right\} \bigcup \left\{\beta_j^{(0)}\right\} \bigcup \left\{\gamma_{ij}^{(0)}\right\}$
        \item A tolerance level $\epsilon$ and a maximum number of iterations .
    \end{itemize}
    
    \textbf{Output:}  
    Estimated model parameters:  
    $\hat{\boldsymbol{\vartheta}} = \{\hat{\theta}_i\} \bigcup \{\hat{\alpha}_j\} \bigcup \{\hat{\beta}_j\} \bigcup \{\hat{\gamma}_{ij}\}$
    
    \begin{algorithmic}[1]     
        \State \textbf{Initialize} parameters $\boldsymbol{\vartheta}^{(0)}$
        \Repeat
            \State \textbf{E-step}: Compute the Q-function described in \eqref{eqn:q_func} based on the current estimates $\boldsymbol{\vartheta}^{(t-1)}$
            \State \textbf{M-step}: Update the parameters by maximizing the Q-function obtained in the E-step. The detailed update rules are provided in \eqref{eqn:update_rule}.
        \Until{The convergence criterion in \eqref{con_cri} is met or the maximum number of iterations is reached.}
    \end{algorithmic}
\end{algorithm}

\section{Theory} \label{sec_4}
In this section, we first justify the use of the prior distribution for $ \gamma_{ij} $ described in \eqref{eqn:prior_gamma} by illustrating its connection to the spike-and-slab prior (Section \ref{sub_sec_4.1}). We then demonstrate that the introduction of the shift parameter $ \gamma_{ij} $ does not exacerbate the identifiability issue presented in one-dimensional BIRT model by \cite{Clinton2004} (Section \ref{sub_sec_4.2}).

\subsection{Spike-and-Slab Prior and Its Connection to $l_0$ Regularization}\label{sub_sec_4.1}
Recall that $ \Gamma \in \mathbb{R}^{I \times J} $ is the matrix composed of $ \gamma_{ij} $. In this paper, we assume that $ \Gamma $ has a sparse structure, meaning that most of its elements are zero, with only a small number of non-zero elements. In other words, most votes are sincere, reflecting the legislator's true utility, while a small number of votes are strategic and do not align with the legislator's true utility.
In order to incorporate the sparse structure of $ \Gamma $ into the model within the Bayesian framework, a natural approach is to use a spike-and-slab prior for $ \gamma_{ij} $:
\begin{equation*}
    \gamma_{ij} \sim (1-\pi) \cdot \delta_0 + \pi \cdot N(0, \sigma_\gamma^2),
\end{equation*}
where $ \delta_0 $ serves as the `spike' component, $ N(0, \sigma_\xi^2) $ serves as the `slab' component, and $ \pi \in (0,1) $ denotes the proportion of the `slab' component, which is related to the level of sparsity of $\Gamma$.

Following \cite{Polson_2018}, we reparameterize $ \gamma_{ij} $ as the product of two random variables:
\begin{align*}
    \gamma_{ij} = \tau_{ij} \cdot \kappa_{ij},
\end{align*}
where $\tau_{ij} \sim \text{Ber}(\pi)$, $\kappa_{ij} \sim N(0, \sigma_\gamma^2)$, and $\tau_{ij} \indep \kappa_{ij}$. The corresponding posterior distribution is then by
\begin{align}\label{eqn:post_spike_slap}
    \begin{split}
        &p\left(\{y_{ij}^*\}, \{\theta_i\}, \{\alpha_j\}, \{\beta_j\}, \{\tau_{ij}\}, \{\kappa_{ij}\} \mid \{y_{ij}\}\right) \\
        &\propto \prod_{i=1}^I \prod_{j = 1}^J \bigg[ \left(\delta_1 (y_{ij})\cdot \mathbb{I}(y_{ij}^* \geq 0) + \delta_0 (y_{ij})\cdot \mathbb{I}(y_{ij}^* < 0) \right) \times  \phi_1\left(y_{ij}^*; \boldsymbol{\tilde{\beta}}_j^\top \boldsymbol{\tilde{\theta}}_i + \tau_{ij}\cdot \kappa_{ij}, 1\right)\\
        &\quad  \times \phi_1\left(\theta_i; \mu_{\theta}, \sigma_{\theta}^2 \right) \times \phi_{2}\left(\boldsymbol{\tilde{\beta}}_j ; \boldsymbol{\mu}_{\boldsymbol{\tilde{\beta}}}, \Sigma_{\boldsymbol{\tilde{\beta}}}\right) \times  (\pi)^{\tau_{ij}}(1-\pi)^{1-\tau_{ij}} \times \phi_1(\kappa_{ij}; 0, \sigma_\gamma^2) \bigg].
    \end{split}
\end{align}
By maximizing the posterior described in \eqref{eqn:post_spike_slap}, we can obtain a MAP estimates for the parameters of interest. Therefore, we are interested in solving the optimization problem outlined in \eqref{eqn:obj_spike_slab}.
\begin{equation}\label{eqn:obj_spike_slab}
    \underset{\theta_i, \boldsymbol{\tilde{\beta}}_j, \tau_{ij}, \kappa_{ij}}{\max} ~\sum_{i=1}^I \sum_{j = 1}^J \bigg[ \zeta_{ij} - \log\left(\dfrac{1-\pi}{\pi}\right) \tau_{ij} - \dfrac{1}{2\sigma_\gamma^2} \kappa_{ij}^2 - \dfrac{1}{2}\left\{ y_{ij}^* - (\boldsymbol{\tilde{\beta}}_j^\top \boldsymbol{\tilde{\theta}}_i + \tau_{ij}\cdot \kappa_{ij})\right\}^2 \bigg],
\end{equation}
where $\zeta_{ij} = \zeta_{ij}(\theta_i, \boldsymbol{\tilde{\beta}}_i) = \log\left(\phi_1\left(\theta_i; \mu_{\theta}, \sigma_{\theta}^2 \right) \cdot \phi_{2}\left(\boldsymbol{\tilde{\beta}}_j ; \boldsymbol{\mu}_{\boldsymbol{\tilde{\beta}}}, \Sigma_{\boldsymbol{\tilde{\beta}}}\right)\right)$.

Theorem~\ref{theorem1} shows that, under its assumptions, finding a MAP estimate that maximizes the posterior distribution described in \eqref{eqn:post_spike_slap} is equivalent to maximizing the posterior distribution in \eqref{eqn:post_l0}. In other words, under these assumptions, obtaining a MAP estimate using the spike-and-slab prior for $\gamma_{ij}$ is equivalent to employing the prior distribution in Equation~\eqref{eqn:prior_gamma}. This equivalence justifies our use of the prior in \eqref{eqn:prior_gamma}.
The proof of Theorem \ref{theorem1} is provided in Appendix \ref{Appen_A.1}.
\begin{theorem}\label{theorem1}
    Assume that $ \pi < \frac{1}{2} $ and $ \sigma_\gamma^2 \to \infty $. For $ \lambda = \left( 2 \log \left( \frac{1-\pi}{\pi} \right) \right)^{1/2} $, solving the optimization problem in \eqref{eqn:obj_spike_slab} is equivalent to solving the optimization problem in \eqref{eqn:obj_l0}, which is derived from \eqref{eqn:post_l0}:
    \begin{equation}\label{eqn:obj_l0}
        \underset{\theta_i, \boldsymbol{\tilde{\beta}}_j, \gamma_{ij}}{\max} ~\sum_{i=1}^I \sum_{j = 1}^J \left[ \zeta_{ij} - \frac{\lambda^2}{2}\mathbb{I}(\gamma_{ij} \neq 0) - \frac{1}{2}\left\{ y_{ij}^* - (\boldsymbol{\tilde{\beta}}_j^\top \boldsymbol{\tilde{\theta}}_i + \gamma_{ij}) \right\}^2 \right],
    \end{equation}
    in the sense that for any arbitrary solution of \eqref{eqn:obj_spike_slab}, we can induce a solution of \eqref{eqn:obj_l0}, and vice versa.
\end{theorem}

The prior in \eqref{eqn:prior_gamma} can be interpreted as imposing $ \ell_0 $-regularization on $\Gamma$ \citep{Polson_2018}. Specifically, taking the logarithm of the prior in \eqref{eqn:prior_gamma} yields
\begin{align*} \sum_{i=1}^I \sum_{j=1}^J -\frac{\lambda^2}{2} \mathbb{I}(\gamma_{ij} \neq 0) = -\frac{\lambda^2}{2} ||\Gamma||_0, \end{align*} where $||A||_0$ denotes the number of nonzero elements in matrix $A$. Thus, Theorem~\ref{theorem1} can be seen as establishing a connection between the spike-and-slab prior and $ \ell_0 $-regularization. This insight motivates the name of our proposed method, \texttt{emRIRT\_L0}, introduced in Section~\ref{sub_sec_5.2}.

\subsection{Identifiability}\label{sub_sec_4.2}
The BIRT model considered by \cite{Clinton2004} and \cite{Imai2016} is unidentifiable. \cite{Gelman2005} classifies the identifiability issues in the Bayesian item response theory model into three types: additive aliasing, multiplicative aliasing, and reflection invariance (sign-flip invariance\footnote{The term `sign-flip invariance' means that the two different parameterizations of BIRT model, 
\begin{align*}
\boldsymbol{\vartheta}^1 = \{\alpha_j\} \cup \{\beta_j\} \cup \{\theta_i\} \quad \text{and}  \quad \boldsymbol{\vartheta}^2 = \{\alpha_j\} \cup \{-\beta_j\} \cup \{-\theta_i\} ,
\end{align*}
yield the same probability of $ Y_{ij} = 1 $ for all $ i = 1, 2, \ldots, I $ and $ j = 1, 2, \ldots, J $. This is clear from the fact that 
\begin{align*}
    \alpha_j + \beta_j \theta_i = \alpha_j + (-\beta_j)(-\theta_i) \quad \text{for all}\quad  i = 1, 2, \ldots, I, \text{ and } j = 1, 2, \ldots, J.
\end{align*}} in the one-dimensional case). In the one-dimensional case, one approach to ensure identifiability up to sign-flip is to impose constraints on the ideal points, such that their mean is 0 and their variance is 1.
Theorem \ref{theorem2} demonstrates that the introduction of the shift parameter $ \gamma_{ij} $ does not exacerbate the identifiability issues inherent in the BIRT model, under the sparsity assumption for $ \Gamma $. The proof of Theorem \ref{theorem2} is provided in Appendix \ref{Appen_A.2}.

\begin{theorem}\label{theorem2}
    Consider a one-dimensional robust BIRT model. Let 
    $\mathcal{I}=\{1,\ldots,I\}$ and $\mathcal{J}=\{1,\ldots,J\}$, and define
    \begin{align*}
        \Delta 
        = \left\{
        \Gamma \in \mathbb{R}^{I \times J} :
        \left| \left\{ 
        i \in \mathcal{I} : \sum_{j \in \mathcal{J}} \mathbb{I}(\gamma_{ij} \neq 0) > 0 
        \right\} \right| \le \frac{I}{2}-1,\;
        \left| \left\{ 
        j \in \mathcal{J} : \sum_{i \in \mathcal{I}} \mathbb{I}(\gamma_{ij} \neq 0) > 0 
        \right\} \right| \le \frac{J}{2}-1
        \right\},
    \end{align*}
    where $|A|$ denotes the cardinality of a set $A$. Assume the following identification conditions:
    \begin{equation*}
        \text{(1) }\Gamma \in \Delta,\qquad
        \text{(2) }\sum_{i\in\mathcal{I}} \theta_i = 0,\qquad
        I^{-1}\sum_{i\in\mathcal{I}} \theta_i^2 = 1.
    \end{equation*}
    Suppose that the parameter sets 
    $(\alpha_{j,1}, \beta_{j,1}, \theta_{i,1}, \gamma_{ij,1})$ and 
    $(\alpha_{j,2}, \beta_{j,2}, \theta_{i,2}, \gamma_{ij,2})$ 
    yield identical joint likelihoods:
    \begin{align}\label{eqn:eqn1}
        \prod_{i\in\mathcal{I}, j\in\mathcal{J}}
        p(y_{ij} \mid \alpha_{j,1}, \beta_{j,1}, \theta_{i,1}, \gamma_{ij,1})
        =
        \prod_{i\in\mathcal{I}, j\in\mathcal{J}}
        p(y_{ij} \mid \alpha_{j,2}, \beta_{j,2}, \theta_{i,2}, \gamma_{ij,2}),
        \qquad 
        \forall\, \{y_{ij}\}\in\{0,1\}^{IJ}.
    \end{align}
    Then exactly one of the cases (A) and (B) holds.
    \begin{align}\label{eqn:eqn2}
        \begin{cases}
        (A)~(\alpha_{j,1},\beta_{j,1},\theta_{i,1})
             =(\alpha_{j,2},\beta_{j,2},\theta_{i,2}),
             &\text{for all }\, i\in\mathcal{I} \text{ and }\ j\in\mathcal{J},\\[4pt]
        (B)~(\alpha_{j,1},\beta_{j,1},\theta_{i,1})
             =(\alpha_{j,2},-\beta_{j,2},-\theta_{i,2}),
             &\text{for all }\, i\in\mathcal{I} \text{ and }\ j\in\mathcal{J}.
        \end{cases}
    \end{align}
\end{theorem}

In Theorem \ref{theorem2}, Assumption (2) is a commonly used condition to ensure the identifiability of the one-dimensional BIRT model up to a sign-flip. Therefore, the only additional assumption needed for identifiability with the inclusion of the shift parameter $ \gamma_{ij} $ is Assumption (1), which imposes a restriction on the class of $ \Gamma $. In many cases, it is reasonable to assume that most votes are sincere, while strategic votes are in the minority. Therefore, Assumption 1 is likely to hold in many practical scenarios.

A limitation of Theorem \ref{theorem2} is that it guarantees identifiability of our proposed BIRT model only up to a sign-flip. However, in the one-dimensional case, the sign of the ideal points can be determined using party information. For example, one can fully identify the ideal points by constraining the most extreme conservative legislators to have positive ideal points.

\section{Simulation Study}\label{sec_5}
\subsection{Roll Call Data Generation}
Following the approach in \cite{Imai2016}, we generate the simulation data using a parametric bootstrap method. Specifically, we first estimate the bill parameters ($ \beta $) and ideal points ($ \theta $) by applying \texttt{emIRT} methods outlined in Section \ref{sub_sec_5.2} to the 112th U.S. Congress roll call data. Legislators with negative estimated ideal points are considered to belong to the Democratic Party, while those with positive estimated ideal points are assigned to the Republican Party. We treat these estimates as the ground truth values and generate the simulation data while preserving the number of bills, and the number of legislators from the pre-processed 112th U.S. Congress roll call data to ensure realistic simulation conditions. The choice of the 112th Congress data is particularly appropriate as it predates the systematic emergence of protest voting behavior, providing a clean baseline for our simulation. 

Using the parametric bootstrap method mentioned above, we can generate roll call data consisting only of sincere votes. To obtain roll call data that includes protest votes, we just replace some of the sincere voting results with the opposite outcome. Specifically, we randomly designate four Democratic legislators with extreme ideal points as protest voters. Next, we randomly select bills with large absolute values of the difficulty parameter ($ |\beta| $), as these bills present the most effective opportunities for protest voting. The number of selected bills depends on the simulation settings. We then invert the voting results corresponding to the selected legislators and bills (i.e., changing yea to nay and nay to yea), thus creating protest votes.

Using the data generation schemes outlined above, for all simulation settings, we have a $ 395 \times 1455 $\footnote{Before applying \texttt{emIRT} methods to the 112th U.S. Congress roll call data, we first perform the standard preprocessing steps outlined in Appendix \ref{Appen_C}. As a result, the number of legislators and bills is reduced compared to the original 112th U.S. Congress roll call data.} roll call matrix, where each row corresponds to a legislator and each column corresponds to a bill. In the case where protest votes are included, there are 4 fixed protest voters, while the number of protest bills varies depending on the simulation settings.

\subsection{Compared Methods}\label{sub_sec_5.2}
We compare three methods :
\begin{itemize}
    \item \texttt{emIRT}: This method uses the posterior distribution described in \eqref{eqn:post_BIRT} as its objective function. A MAP estimate for the parameter of interest is obtained by applying the EM algorithm introduced by \cite{Imai2016}.

    \item \texttt{emRIRT\_L0}: This method, proposed in this paper, uses the posterior distribution described in \eqref{eqn:post_l0} as its objective function, and a MAP estimate is obtained by applying the EM algorithm outlined in Section \ref{sub_sec_3.2}. The third letter, `\texttt{R}', in $\texttt{emRIRT\_L0}$ stands for `robust', highlighting the model's robustness against strategic votes. The term `\texttt{L0}' refers to the type of prior used for the shift parameter $ \gamma_{ij} $, specifically indicating $ \ell_0 $-regularization for $ \Gamma $.

    \item \texttt{emRIRT\_L1}: This method is identical to \texttt{emRIRT\_L0} except for the prior distribution of the shift parameter $ \gamma_{ij} $. Instead of the prior in \eqref{eqn:prior_gamma}, \texttt{emRIRT\_L1} uses a Laplace prior given by $p(\gamma_{ij} \mid \lambda) \propto \exp\left(-\lambda|\gamma_{ij}|\right)$,
    which is equivalent to applying $ \ell_1 $-regularization to $ \Gamma $. The term `\texttt{L1}' reflects this change. After deriving an EM algorithm under this prior distribution, we apply it to obtain a MAP estimate.
\end{itemize}

\subsection{Results}

\begin{figure}[!htb]
    \centering
    \includegraphics[width= 0.8\linewidth]{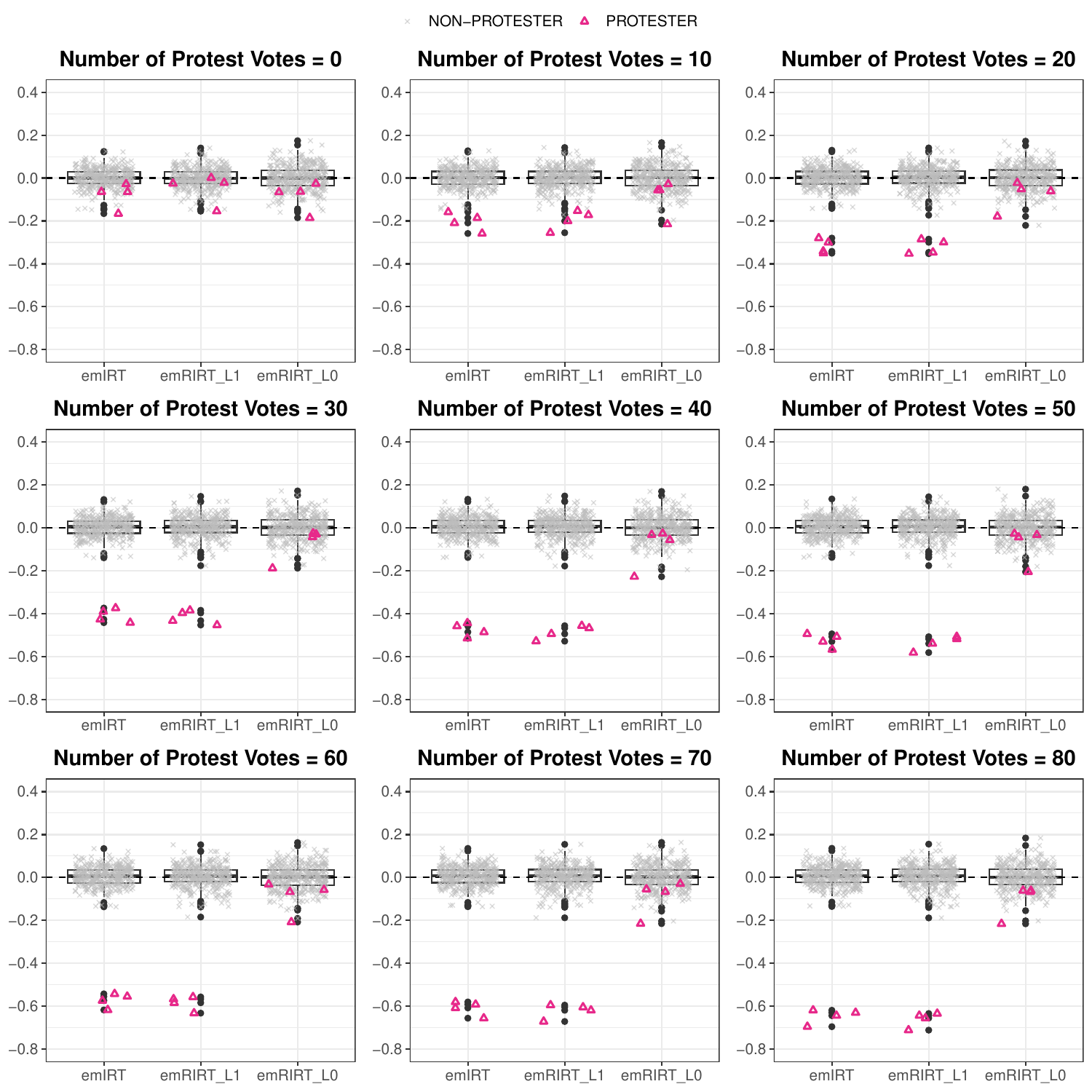}
    \caption{Boxplot of estimation errors in ideal points ($\theta_i^* - \hat{\theta_i}$) across varying numbers of protest votes. Each panel displays results for a specified number of protest votes. The $x$-axis represents the method used to estimate ideal points, while the $y$-axis shows the estimation error (true ideal point minus estimated ideal point). Grey crosses indicate the estimation errors for individual legislators, while pink triangles highlight the estimation errors for the four protest legislators.}
    \label{fig:shrinkage}
\end{figure}

Figure~\ref{fig:shrinkage} presents simulation results for settings in which each of the four protest legislators casts between 0 and 80 protest votes, in increments of 10. For example, a panel labeled ``Number of Protest Votes = 80" indicates that each of the four protest legislators casts 80 protest votes, so the dataset contains a total of $4 \times 80 = 320$ protest votes.
Across all levels of protest voting, both \texttt{emIRT} and \texttt{emRIRT\_L1} yield estimation errors for the protest legislators (pink triangles) that remain far from zero. These errors become increasingly negative as the number of protest votes grows, which indicates a pronounced attenuation bias. In contrast, our proposed method (\texttt{emRIRT\_L0}) keeps the errors consistently close to zero across all settings, demonstrating robustness to varying degrees of protest behavior.
In the extreme case where each of the four protest legislators casts 80 protest votes, the estimation errors for the protest legislators from \texttt{emIRT} and \texttt{emRIRT\_L1} fall below $-0.6$, whereas the corresponding errors produced by \texttt{emRIRT\_L0} remain within $[-0.2,\, 0.2]$.

\begin{figure}[!htb]
    \centering
    \includegraphics[width=0.9\linewidth]{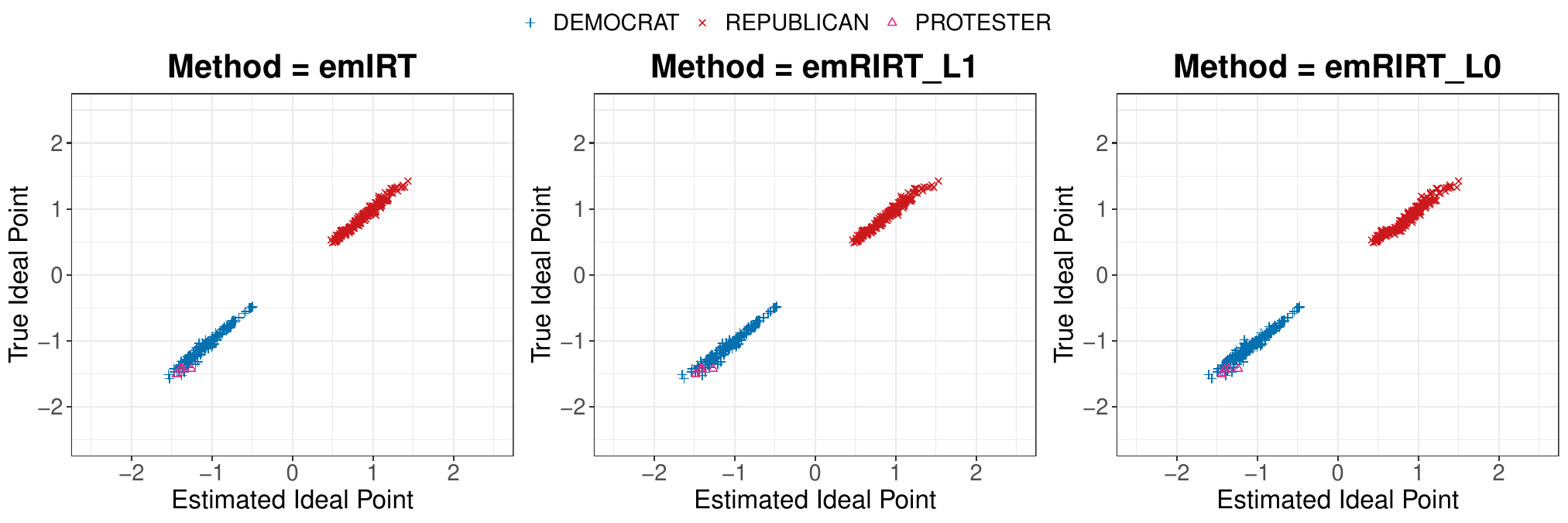}
 \caption{Comparison of ideal point estimates without protest votes. The results show that the choice of \texttt{emIRT}, \texttt{emRIRT\_L1}, or \texttt{emRIRT\_L0} does not affect estimates in the absence of protest votes. 
        Each panel plots estimates ($x$-axis) against the ground truth ($y$-axis). 
        The left panel shows \texttt{emIRT} estimates, 
        the center panel presents \texttt{emRIRT\_L1} estimates, and 
        the right panel displays \texttt{emRIRT\_L0} estimates. 
        Blue pluses represent Democrats, and red crosses denote Republicans.}   \label{fig:simulno}
\end{figure}

\begin{figure}[!htb]
    \centering
    \includegraphics[width=0.9\linewidth]{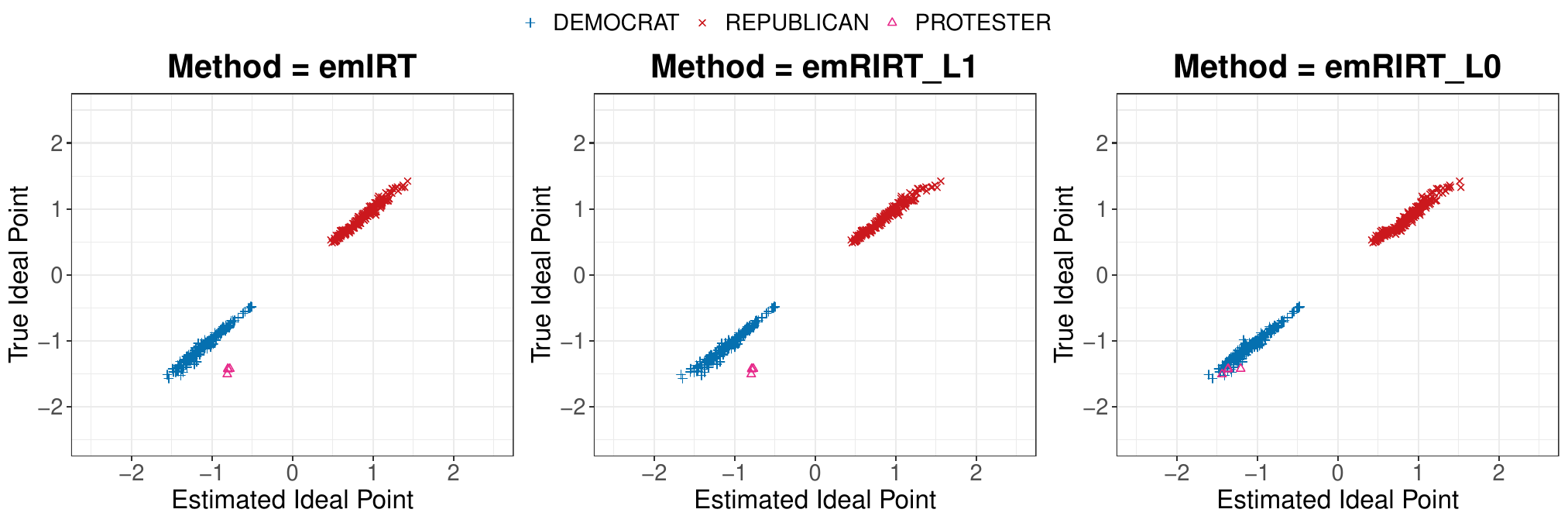}
    \caption{Comparison of ideal point estimates when each of the four protest legislators casts 80 protest votes. The results show that the use of \texttt{emIRT} or \texttt{emRIRT\_L1} produces attenuated estimates for ideal point estimates of protest legislators while \texttt{emRIRT\_L0} does not.      
    Each panel plots estimates ($x$-axis) against the ground truth ($y$-axis). 
        The left panel shows \texttt{emIRT} estimates, 
        the center panel presents \texttt{emRIRT\_L1} estimates, and 
        the right panel displays \texttt{emRIRT\_L0} estimates. 
        Blue pluses represent Democrats, and red crosses denote Republicans. Triangles indicate protest legislators.}
    \label{fig:simulpro}
\end{figure}

Figure \ref{fig:simulno} compares ideal point estimates from three methods (\texttt{emIRT}, \texttt{emRIRT\_L1}, and \texttt{emRIRT\_L0}) in the absence of protest votes. The estimates from all three methods are highly consistent, with Democrats (blue pluses) and Republicans (red crosses) showing similar ideological positions across methods. 
Figure~\ref{fig:simulpro} presents the same comparison when each of the four protest legislators casts 80 protest votes. In this case, the estimates diverge notably for protest legislators (indicated by triangles). Both \texttt{emIRT} and \texttt{emRIRT\_L1} show significant attenuation bias in their estimates, while our proposed method, \texttt{emRIRT\_L0}, maintains robust estimates that better reflect these legislators' true ideological positions.

The non-robustness of the L1 penalty in robust regression has been extensively discussed in \cite{she2011}, and we therefore do not elaborate on it here but simply refer the reader to their paper.

\section{Application : 116th and 117th House of Representatives}\label{sec_6}

We evaluate our proposed method, \texttt{emRIRT\_L0}, by analyzing roll-call votes from the 116th (2019-2020) and 117th (2021-2022) U.S. House of Representatives and comparing results with the conventional \texttt{emIRT} approach. For both congresses, we compiled complete voting records from official House sources. We implemented standard preprocessing procedures, which are detailed in Appendix \ref{Appen_C}. The validity assessment of strategic voting outcomes detected by \texttt{emRIRT\_L0} for the 116th House of Representatives, as well as the application of the method to the 118th House, is provided in Appendix~\ref{Appen_D} and Appendix~\ref{Appen_E}, respectively.

 \subsection{Changes in Ideal Point Estimates}\label{sec_61}

\begin{figure}[!htb]
\centering
\begin{overpic}[width=0.4\linewidth]{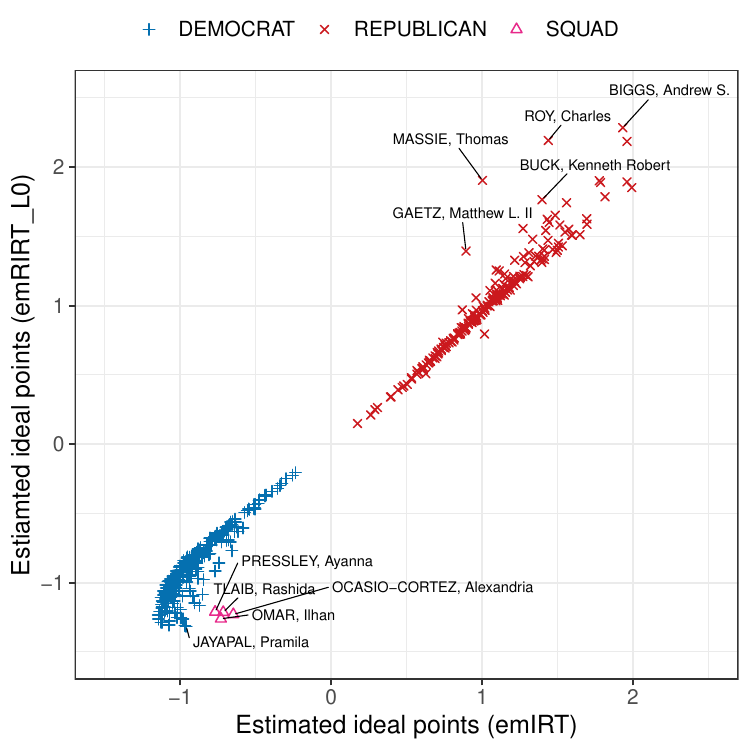}
    \put(2,95){\textbf{(A)}}
\end{overpic}
\begin{overpic}[width=0.4\linewidth]{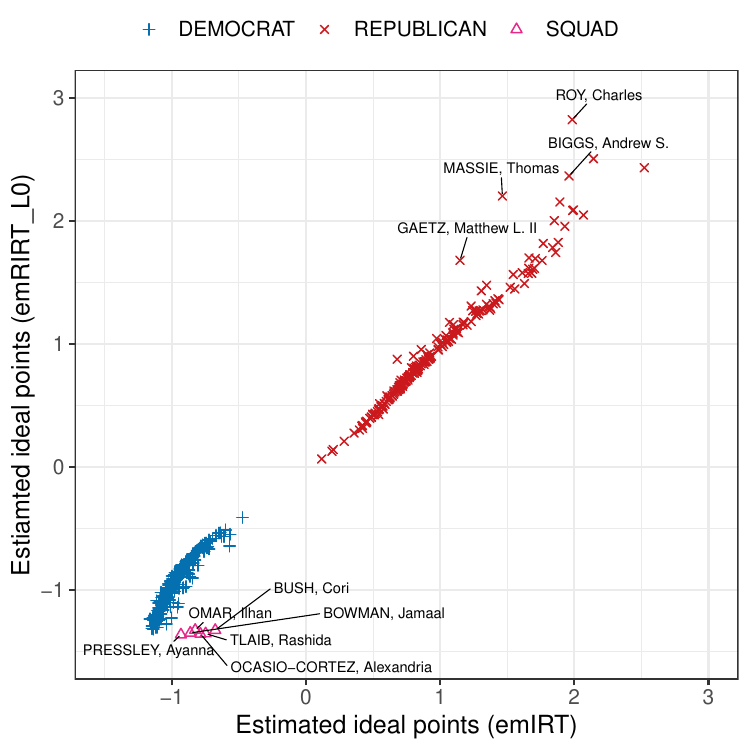}
    \put(2,95){\textbf{(B)}}
\end{overpic}
\caption{Comparison of ideal point estimates for the 116th and 117th Congress. In panel A (116th Congress) and panel B (117th Congress), the $x$-axis represents the ideal point estimates via \texttt{emIRT}, while the $y$-axis shows the estimates from \texttt{emRIRT\_L0}. Democratic members are indicated by blue pluses, Republican members by red crosses, and the Squad members by pink triangles.}
\label{fig:H1167}
\end{figure}

Figure \ref{fig:H1167} compares the ideal point estimates obtained from \texttt{emIRT} ($x$-axis) and \texttt{emRIRT\_L0} ($y$-axis) for both the 116th (penal A) and 117th (panel B) U.S. House of Representatives. While the estimates show strong linear correlation for most legislators (appearing along the diagonal), we observe substantial deviations for several members at both ideological extremes.

In the 116th House (penal A), the Squad members (represented by pink triangles) exhibit the most dramatic shifts among Democrats (blue pluses). Specifically, Alexandria Ocasio-Cortez, Ilhan Omar, Rashida Tlaib, and Ayanna Pressley appear moderately liberal under \texttt{emIRT} (with $x$-values around -1) but are identified as far more progressive under \texttt{emRIRT\_L0} (with $y$-values below -1.2). Among Republicans (red crosses), significant rightward shifts are observed for Freedom Caucus and libertarian-leaning members- most notably Thomas Massie, Andrew S. Biggs, Charles Roy, Kenneth Robert Buck, and Matthew L. Gaetz - who cluster distinctly above the main diagonal with \texttt{emRIRT\_L0} estimates exceeding 1.5, compared to their more moderate \texttt{emIRT} positions.

The 117th House (panel B) displays a similar pattern with even more pronounced shifts. The expanded Squad (now including Cori Bush and Jamaal Bowman alongside the original four members) shows extreme liberal positioning under \texttt{emRIRT\_L0} ($y$-values around -1.3) despite appearing less extreme in conventional estimation. On the Republican side, Charles Roy, Andrew S. Biggs, Thomas Massie, and Matthew L. Gaetz again stand out with \texttt{emRIRT\_L0} estimates placing them substantially further to the right ($y$-values above 2.0) than \texttt{emIRT} estimates would suggest.

In both congresses, \texttt{emIRT} systematically produces more moderate estimates than \texttt{emRIRT\_L0} for legislators who engage in protest voting, effectively compressing the ideological scale and obscuring the true extent of polarization. 

\begin{table}[!htb]
\small
\centering
\caption{Largest Changes in Ideological Rankings between Methods for the 116th Congress: Values in the fifth and sixth columns show empirical quantiles of estimated ideal points (0 $\approx$ most liberal, 1 = most conservative). $|\Delta|$ represents the absolute difference between \texttt{emIRT} and  \texttt{emRIRT\_L0} quantiles.}
\begin{tabular}{llccrrr}
\toprule
\textbf{Representative} & \textbf{District} & \textbf{Party} & \textbf{Squad} & \texttt{emIRT} & \texttt{emRIRT\_L0} & \textbf{$|\Delta|$} \\ 
\midrule
116the Congress\\
\midrule
Alexandria Ocasio-Cortez & NY-14 & D & Yes & 0.475 & 0.032 & 0.443 \\ 
Ilhan Omar               & MN-5  & D & Yes & 0.430 & 0.020 & 0.410 \\ 
Rashida Tlaib            & MI-13 & D & Yes & 0.439 & 0.045 & 0.394 \\ 
Ayanna Pressley          & MA-7  & D & Yes & 0.405 & 0.043 & 0.362 \\ 
Peter DeFazio            & OR-4  & D & No  & 0.345 & 0.068 & 0.277 \\ 
Pramila Jayapal          & WA-7  & D & No  & 0.245 & 0.002 & 0.243 \\ 
Sylvia Garcia            & TX-29 & D & No  & 0.356 & 0.119 & 0.237 \\ 
Thomas Massie            & KY-4  & R & No  & 0.755 & 0.989 & 0.234 \\ 
Mark Pocan               & WI-2  & D & No  & 0.236 & 0.005 & 0.231 \\ 
Matt Gaetz               & FL-1  & R & No  & 0.696 & 0.926 & 0.230 \\ 
Ro Khanna                & CA-17 & D & No  & 0.306 & 0.077 & 0.229 \\ 
\midrule
117the Congress\\
\midrule
Cori Bush               & MO-1  & D & Yes  & 0.486 & 0.005 & 0.481 \\
Rashida Tlaib           & MI-13 & D & Yes  & 0.459 & 0.007 & 0.452 \\
Alexandria Ocasio-Cortez & NY-14 & D & Yes  & 0.430 & 0.009 & 0.421 \\
Ilhan Omar              & MN-5  & D & Yes  & 0.408 & 0.016 & 0.392 \\
Jamaal Bowman           & NY-16 & D & Yes  & 0.378 & 0.011 & 0.367 \\
Ayanna Pressley         & MA-7  & D & Yes  & 0.288 & 0.002 & 0.286 \\
Maxine Waters           & CA-43 & D & No   & 0.250 & 0.072 & 0.178 \\
Jefferson Van Drew      & NJ-2  & R & No   & 0.617 & 0.784 & 0.167 \\
Veronica Escobar        & TX-16 & D & No   & 0.255 & 0.106 & 0.149 \\
Mark Pocan              & WI-2  & D & No   & 0.187 & 0.047 & 0.140 \\
Rosa L. DeLauro         & CT-3  & D & No   & 0.387 & 0.248 & 0.139 \\
\bottomrule
\end{tabular}\label{tab:legislatordata}
\end{table}

Table \ref{tab:legislatordata} provides evidence of how accounting for protest votes fundamentally alters our understanding of legislator ideology. The quantile differences between \texttt{emIRT} and our proposed \texttt{emRIRT\_L0} reveal systematic patterns of misrepresentation of ideological spectrum that has significant implications for the study of the U.S. Congress. 

The most striking finding is the substantial repositioning of Squad members in both congresses. In the 116th Congress, Alexandria Ocasio-Cortez exhibits the largest quantile shift of any legislator (0.443), moving from a surprisingly moderate position (0.475) under conventional estimation to a firmly progressive one (0.032) when accounting for protest votes. This finding is remarkable because conventional methods place her near the median of the Democratic caucus, which contradicts her well-documented progressive policy positions and public persona.

Similarly, Ilhan Omar ($|\Delta|$ = 0.410), Rashida Tlaib ($|\Delta|$ = 0.394), and Ayanna Pressley ($|\Delta|$ = 0.362) all experience shifts exceeding 0.36 in quantile space. These adjustments are not merely statistical artifacts but reflect meaningful corrections that better align with these legislators' self-identified progressive positioning and observable voting patterns.

The pattern becomes even more pronounced in the 117th Congress, where Cori Bush shows an extraordinary shift of 0.481 quantile points-the largest in our dataset. All six Squad members demonstrate quantile shifts exceeding 0.28, with five of them repositioned to extremely progressive positions (quantiles below 0.02). This consistency across both congresses validates our method's ability to detect and adjust for protest voting behavior.

Beyond the Squad, our analysis reveals significant quantile shifts for other progressive Democrats. Progressive Caucus Chair Pramila Jayapal shows a remarkable shift of 0.243 in the 116th Congress, moving from a moderate-progressive position (0.245) to the extreme progressive end of the spectrum (0.002). Similarly, Mark Pocan, another Progressive Caucus leader, moves 0.231 quantile points leftward.

These corrections suggest that conventional methods have systematically underestimated the progressive positioning of the Democratic left wing. Under \texttt{emIRT}, these legislators appear as mainstream Democrats, but \texttt{emRIRT\_L0} reveals them as ideological anchors of the progressive movement, which better matches their policy advocacy and public positioning.

On the Republican side, our method identifies substantial rightward shifts for members associated with the Freedom Caucus and libertarian positions. Thomas Massie, known for his principled libertarian stances and nicknamed ``Mr. No'' for his frequent opposition to spending bills, shifts 0.234 quantile points rightward. \texttt{emRIRT\_L0} places him at 0.989 - essentially the most conservative member of the House - compared to his more moderate 0.755 position under conventional estimation.

Similarly, Matt Gaetz shows a quantile shift of 0.230, moving from 0.696 to 0.926. This correction better reflects his alignment with the Freedom Caucus and his reputation for taking hardline conservative positions while occasionally breaking with party leadership on procedural votes and strategic issues.

The magnitude of these quantile shifts - many exceeding 0.3 or even 0.4 - underscores the significant limitations of conventional ideal point estimation in the presence of legislators' strategic voting behaviors. Traditional methods systematically misplace legislators who engage in protest voting, creating a distorted picture of congressional ideology.

\section{Multidimensional Extension}\label{sec_7}
In this section, we consider a multidimensional extension of our proposed method. First, we generalize the one-dimensional robust BIRT model introduced in Section \ref{sec_3} to a $ K $-dimensional model (Section \ref{sub_sec_7.1}), and present the EM algorithm for the $K$-dimensional robust BIRT model (Section \ref{sub_sec_7.2}). Next, we extend the one-dimensional theory described in Section \ref{sec_4} to its $ K $-dimensional counterpart (Section \ref{sub_sec_7.3}). Finally, focusing on the two-dimensional case, we apply our proposed method, \texttt{emRIRT\_L0}, to roll call data from the 116th and 117th U.S. House of Representatives and compare the results with those obtained using \texttt{emIRT} proposed by \cite{Imai2016} (Section \ref{sub_sec_7.4}).

\subsection{Multidimensional extension of the robust BIRT model}\label{sub_sec_7.1}
We begin with generalizing the definition of the utility function in \eqref{eqn:utility_yea_nay} to its $K$-dimensional counterpart. The utility of voting yea ($U_{ij}^{yea}$) and nay ($U_{ij}^{nay}$) are defined as follows:
\begin{align*}
\begin{split}
    U_{ij}^{\text{yea}} &= U^{\text{yea}}(\boldsymbol{\theta}_i, \boldsymbol{\zeta}_j) = -||\boldsymbol{\theta}_i - \boldsymbol{\zeta}_j||_{L_2}^2 + \eta_{ij},\\
    U_{ij}^{\text{nay}} &= U^{\text{nay}}(\boldsymbol{\theta}_i, \boldsymbol{\psi}_j) = -||\boldsymbol{\theta}_i - \boldsymbol{\psi}_j||_{L_2}^2 + \nu_{ij},
\end{split}
\end{align*}  
where $ \boldsymbol{\theta}_i \in \mathbb{R}^K $ represents legislator $ i $'s ideal point in the policy space, while $ \boldsymbol{\zeta}_j \in \mathbb{R}^K$ and $ \boldsymbol{\psi}_j \in \mathbb{R}^K$ represent the policy positions associated with yea and nay votes for bill $ j $, respectively. The terms $ \eta_{ij} $ and $ \nu_{ij} $ are stochastic errors, assumed to satisfy $ \eta_{ij} - \nu_{ij} \sim N(0, \sigma_j^2) $.   
The utility of legislator $i$ for bill $j$ is then given by  
\begin{align*}
    U_{ij} = U(\boldsymbol{\theta}_i, \boldsymbol{\zeta}_j, \boldsymbol{\psi}_j)  = U_{ij}^{\text{yea}} - U_{ij}^{\text{nay}} = -||\boldsymbol{\theta}_i - \boldsymbol{\zeta}_j||_{L_2}^2 + ||\boldsymbol{\theta}_i - \boldsymbol{\psi}_j||_{L_2}^2 + \xi_{ij} + (\eta_{ij} - \nu_{ij}).
\end{align*}  
By the normality assumption for stochastic error terms, we have
\begin{align*}
    p(U_{ij} \geq 0 \mid \boldsymbol{\theta}_i, \boldsymbol{\zeta}_j, \boldsymbol{\psi}_j, \sigma_j, \xi_{ij}) 
    &= \Phi(\alpha_j + \boldsymbol{\beta}_j^\top \boldsymbol{\theta}_i + \gamma_{ij}),
\end{align*}  
where $ Z_{ij} $ denotes a standard normal random variable, $ \alpha_j = (\boldsymbol{\psi}_j^\top \boldsymbol{\psi}_j - \boldsymbol{\zeta}_j^\top \boldsymbol{\zeta}_j)/\sigma_j $ represents the difficulty parameter, $ \boldsymbol{\beta}_j = 2(\boldsymbol{\zeta}_j - \boldsymbol{\psi}_j)/\sigma_j $ is the discrimination parameter, and $\gamma_{ij} = \xi_{ij}/\sigma_j$ represents the shift parameter.
Analogous to the one-dimensional case, the likelihood for the observed votes is given by
\begin{align}\label{eqn:like_gamma_MD}
    \prod_{i=1}^I \prod_{j=1}^J p(y_{ij} = 1 \mid \alpha_j, \boldsymbol{\beta}_j, \boldsymbol{\theta}_i, \gamma_{ij}) = \prod_{i=1}^I \prod_{j = 1}^J \Phi(\alpha_j + \boldsymbol{\beta}_j^\top \boldsymbol{\theta}_i + \gamma_{ij})^{y_{ij}} \cdot \{1-\Phi(\alpha_j + \boldsymbol{\beta}_j^\top \boldsymbol{\theta}_i + \gamma_{ij})\}^{1-y_{ij}}.
\end{align}  
We adopt Gaussian assumptions for the priors of $\tilde{\boldsymbol{\beta}}_j$ and $\boldsymbol{\theta}_i$ and consider the prior described in \eqref{eqn:prior_gamma} for $\gamma_{ij}$. 
Applying the data-augmented scheme of \cite{AlbertChib1993}, the posterior distribution is given by  
\begin{align}\label{eqn:post_l0_MD}
    \begin{split}
        &p\left(\{y_{ij}^*\}, \{\boldsymbol{\theta}_i\}, \{\alpha_j\}, \{\boldsymbol{\beta}_j\}, \{\gamma_{ij}\} \mid \{y_{ij}\}\right) \\
        &\propto \prod_{i=1}^I \prod_{j = 1}^J 
        \bigg[\left(\delta_0(y_{ij})  \cdot \mathbb{I}(y_{ij}^* \geq 0) + \delta_1(y_{ij}) \cdot \mathbb{I}(y_{ij}^* < 0) \right) \times \phi_1\left(y_{ij}^*; \tilde{\boldsymbol{\beta}}_j^\top \tilde{\boldsymbol{\theta}}_i + \gamma_{ij}, 1\right)\\
        &\quad \quad \times \phi_K\left(\boldsymbol{\theta}_i; \mu_{\boldsymbol{\theta}}, \Sigma_{\boldsymbol{\theta}} \right) \times \phi_{K+1}\left(\tilde{\boldsymbol{\beta}}_j ; \mu_{\tilde{\boldsymbol{\beta}}}, \Sigma_{\tilde{\boldsymbol{\beta}}}\right) \times \exp\left( -\frac{\lambda^2}{2} \mathbb{I}(\gamma_{ij} \neq 0)\right)\bigg],
    \end{split}
\end{align}
where $\tilde{\boldsymbol{\beta}}_j = (\alpha_j, \boldsymbol{\beta}_j^\top)^\top \in \mathbb{R}^{K+1}$ and $\tilde{\boldsymbol{\theta}}_i = (1, \boldsymbol{\theta}_i^\top)^\top \in \mathbb{R}^{K+1}$.

\subsection{EM Algorithm for the multidimensional robust BIRT Model}\label{sub_sec_7.2}
We maximize the posterior in \eqref{eqn:post_l0_MD} using the EM algorithm to obtain a MAP estimate.
The E-step and M-step from the one-dimensional case can be extended to the multidimensional case in a straightforward manner. Therefore, we present only the update rules at the $t$-th iteration.  
Let $ \boldsymbol{\theta}^{(t-1)}, \alpha_j^{(t-1)}, \boldsymbol{\beta}_j^{(t-1)}, \gamma_{ij}^{(t-1)} $ denote the parameter values obtained at iteration $ t-1 $. The closed-form sequential update rule is given by
\begin{align*}
    \boldsymbol{\theta}_i^{(t)} 
    &= \left(\Sigma_{\boldsymbol{\theta}}^{-1} + \sum_{j = 1}^J \boldsymbol{\beta}_j^{(t-1)} \boldsymbol{\beta}_j^{(t-1)\top} \right)^{-1}  \left(\Sigma_{\boldsymbol{\theta}}^{-1} \mu_{\boldsymbol{\theta}} + \sum_{j = 1}^J \boldsymbol{\beta}_j^{(t-1)} \left(y_{ij}^{*(t)} - \gamma_{ij}^{(t-1)} - \alpha_j^{(t-1)}\right)\right), \\ 
    \tilde{\boldsymbol{\beta}}_j^{(t)} 
    &= \left(\Sigma_{\tilde{\boldsymbol{\beta}}}^{-1} + \sum_{i = 1}^I \tilde{\boldsymbol{\theta}}_i^{(t)} \tilde{\boldsymbol{\theta}}_i^{(t)\top}\right)^{-1}  \left( \Sigma_{\tilde{\boldsymbol{\beta}}}^{-1} \mu_{\tilde{\boldsymbol{\beta}}} + \sum_{i = 1}^I \tilde{\boldsymbol{\theta}}_i^{(t)} \left(y_{ij}^{*(t)} - \gamma_{ij}^{(t-1)}\right)\right), \\ 
    \gamma_{ij}^{(t)} 
    &= \left( y_{ij}^{*(t)} - \tilde{\boldsymbol{\beta}}_j^{(t)\top} \tilde{\boldsymbol{\theta}}_i^{(t)}\right) ~ \mathbb{I}\left(~\Big|~y_{ij}^{*(t)} - \tilde{\boldsymbol{\beta}}_j^{(t)\top}\tilde{\boldsymbol{\theta}}_i^{(t)}~\Big|~ > \lambda\right),
\end{align*}
where $ y_{ij}^{*(t)} $ is defined similarly as in \eqref{eqn:y_star}.

\subsection{Multidimensional Extension of Theorem \ref{theorem2}}\label{sub_sec_7.3}
In this section, we extend Theorem \ref{theorem2} from the one-dimensional case to its multidimensional version\footnote{The proof of Theorem \ref{theorem1} does not depend on the dimensionality of $\boldsymbol{\theta}$ and $\boldsymbol{\beta}$. Hence, we can generalize its statement and proof to the multidimensional case in a straightforward manner. Consequently, in this section, we focus exclusively on the multidimensional extension of Theorem \ref{theorem2}.}. Theorem \ref{theorem3} shows that, in a $ K $-dimensional robust BIRT model, the ideal points are identifiable up to an orthogonal transformation under its given assumptions. The proof of Theorem \ref{theorem3} is provided in Appendix \ref{Appen_A.3}.

\begin{theorem}\label{theorem3}
Consider a $K$-dimensional robust BIRT model, where $2 \le K \le \min(I,J)-1$. Let $\mathcal{I}=\{1,\ldots,I\}$ and  $\mathcal{J}=\{1,\ldots,J\}$, and define
\begin{align*}
    \Delta = \left\{
    \Gamma \in \mathbb{R}^{I \times J} :
    \left|\left\{
    i \in \mathcal{I} : \sum_{j\in\mathcal{J}} \mathbb{I}(\gamma_{ij}\neq 0)>0
    \right\}\right| \le \frac{I-K-1}{2},\;
    \left|\left\{
    j \in \mathcal{J} : \sum_{i\in\mathcal{I}} \mathbb{I}(\gamma_{ij}\neq 0)>0
    \right\}\right| \le \frac{J-K-1}{2}
    \right\}.
\end{align*}
Assume the following identification conditions:
\begin{equation*}
    \text{(1) }\Gamma\in\Delta,\qquad
    \text{(2) } \Theta^\top \boldsymbol{1}_{I\times 1} = \boldsymbol{0}_{K\times 1},\qquad
    I^{-1}\Theta^\top \Theta = I_{K\times K},    
\end{equation*}
where $\Theta$ is the $I\times K$ matrix whose $i$-th row equals the ideal point $\boldsymbol{\theta}_i$ of legislator $i$. The vectors $\boldsymbol{1}_{m\times 1}$ and $\boldsymbol{0}_{m\times 1}$ denote the $m$-dimensional vectors of ones and zeros, respectively, and $I_{m\times m}$ denotes the $m\times m$ identity matrix.
Suppose that the parameter sets $(\alpha_{j,1},\boldsymbol{\beta}_{j,1},\boldsymbol{\theta}_{i,1},\gamma_{ij,1})$ and $(\alpha_{j,2},\boldsymbol{\beta}_{j,2},\boldsymbol{\theta}_{i,2},\gamma_{ij,2})$ produce identical joint likelihoods:
\begin{align}\label{eqn:eqn1_MD}
    \prod_{i\in\mathcal{I}, j\in\mathcal{J}}
    p(y_{ij}\mid \alpha_{j,1}, \boldsymbol{\beta}_{j,1}, 
    \boldsymbol{\theta}_{i,1}, \gamma_{ij,1})
    =
    \prod_{i\in\mathcal{I}, j\in\mathcal{J}}
    p(y_{ij}\mid \alpha_{j,2}, \boldsymbol{\beta}_{j,2}, 
    \boldsymbol{\theta}_{i,2}, \gamma_{ij,2}),
    \qquad
    \forall \{y_{ij}\}\in\{0,1\}^{IJ}.
\end{align}
Then there exists an orthogonal matrix $O\in\mathbb{R}^{K\times K}$ such that
\begin{align}\label{eqn:eqn2_MD}
    (\alpha_{j,1}, \boldsymbol{\beta}_{j,1}^\top, \boldsymbol{\theta}_{i,1}^\top)
    =
    (\alpha_{j,2}, (O\boldsymbol{\beta}_{j,2})^\top, 
    (O\boldsymbol{\theta}_{i,2})^\top),
    \qquad
    \text{for all } i\in\mathcal{I} \text{ and }\ j\in\mathcal{J}.
\end{align}
\end{theorem}

\subsection{Application of two-dimensional \texttt{emRIRT\_L0} and \texttt{emIRT}: 116th and 117th House of Representatives}\label{sub_sec_7.4}

\begin{figure}[!hbt]
    \centering

    \begin{overpic}[width=0.8\linewidth]{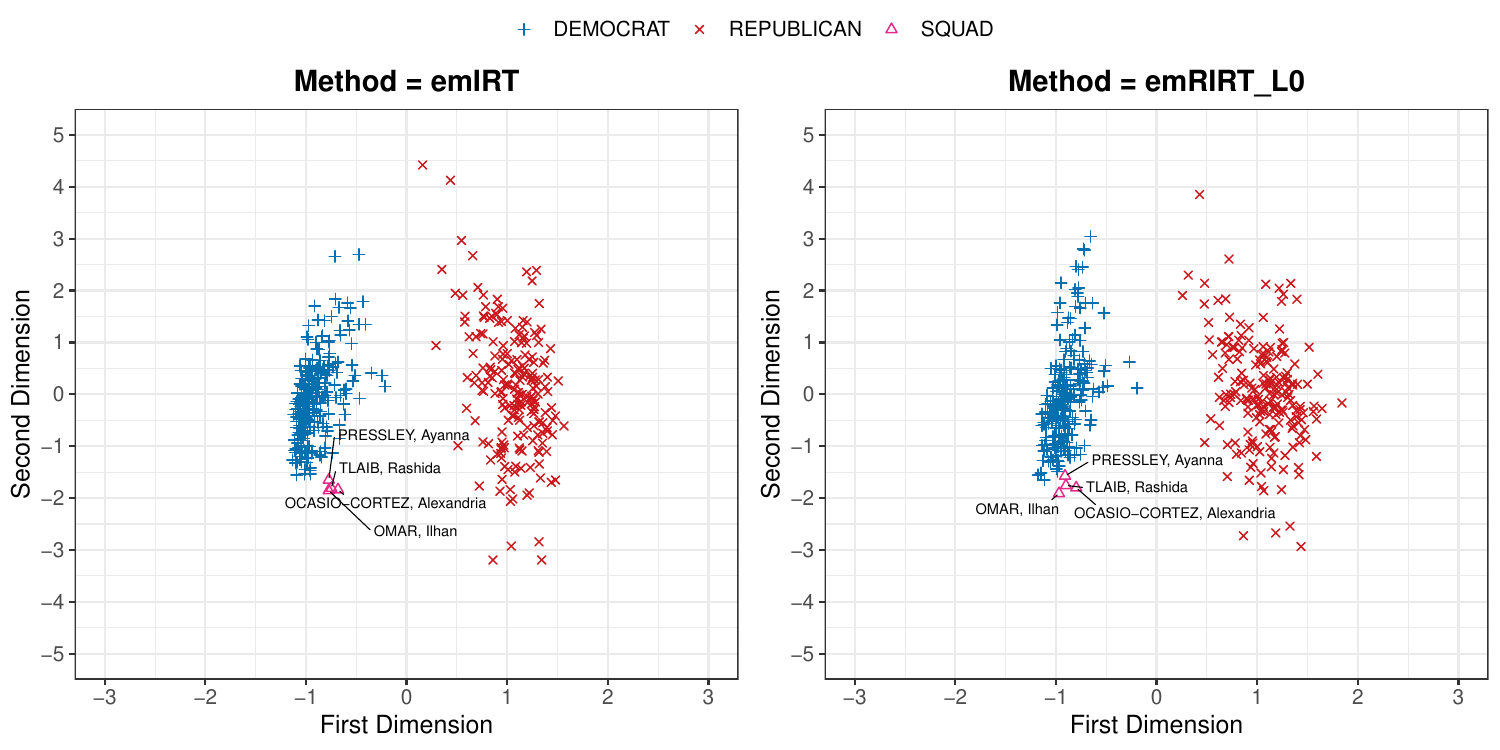}
        \put(2,47){\textbf{(A)}}  
    \end{overpic}
    \vspace{0.5cm}
    \begin{overpic}[width=0.8\linewidth]{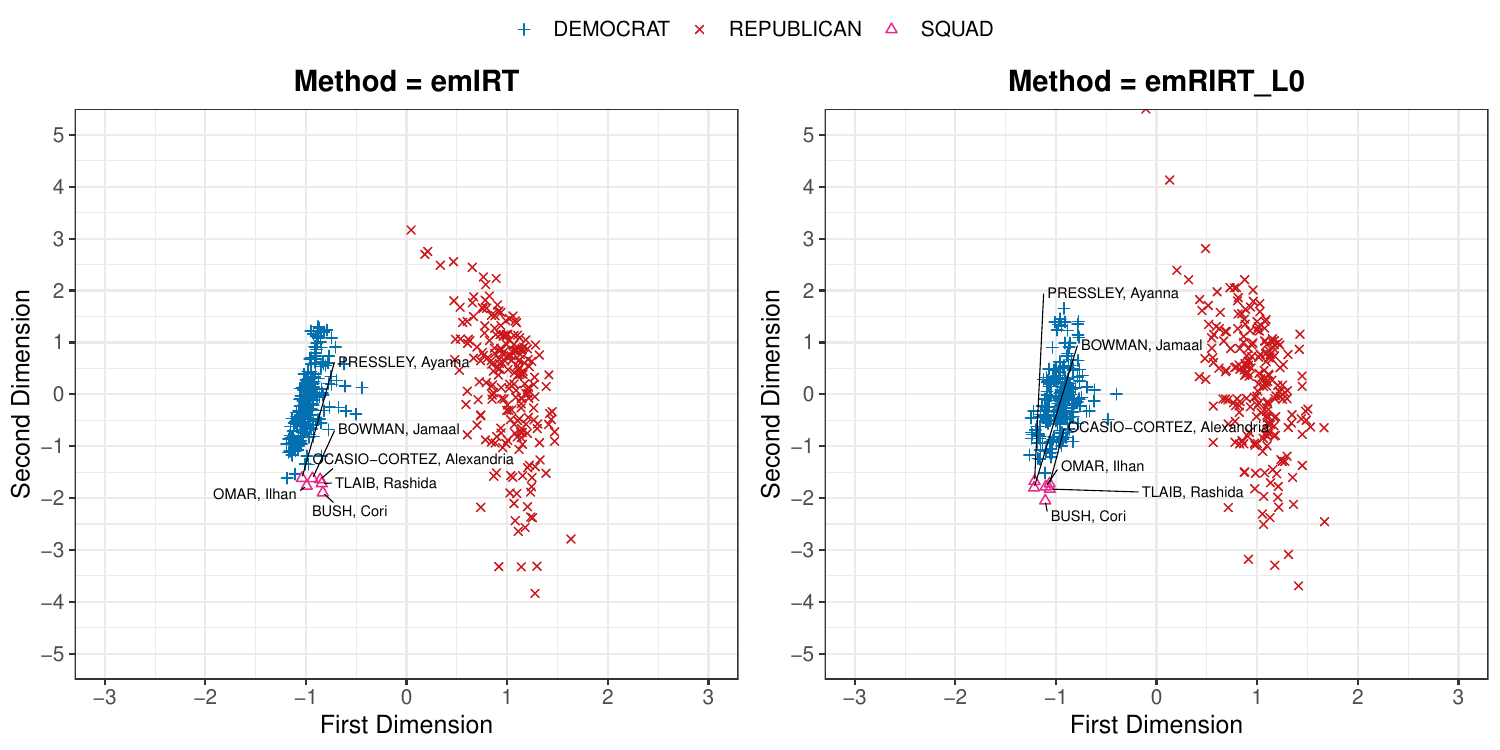}
        \put(2,47){\textbf{(B)}}  
    \end{overpic}

    \caption{Comparison of ideal point estimates for the 116th Congress (panel A) and the 117th Congress (panel B). In each panel, the left plot presents estimates from \texttt{emIRT}, while the right plot shows results from \texttt{emRIRT\_L0}. In all plots, the $x$-axis represents the first dimension of the policy space, and the $y$-axis represents the second dimension. Democratic members are marked with blue plus signs, Republican members with red crosses, and Squad members are highlighted with pink triangles.}
    \label{fig:2d_comparison_1167}
\end{figure}

We extended our analysis to a two-dimensional framework, comparing both the conventional \texttt{emIRT} and our proposed \texttt{emRIRT\_L0} methods in estimating legislators' ideal points in a two-dimensional space. Figure \ref{fig:2d_comparison_1167} presents the two-dimensional ideal point estimates for the 116th (penal A) and 117th (penal B) U.S. House of Representatives under both methods, with Squad members highlighted with pink triangles for easy comparison.\footnote{Application to the 118th House of Representatives is available in Appendix \ref{Appen_E}.}

The results reveal several key patterns consistent with our one-dimensional findings while introducing important nuances. In both congresses, Squad members' ideal points are estimated as more ideologically extreme under our \texttt{emRIRT\_L0} method compared to conventional \texttt{emIRT}. This divergence is particularly pronounced in the 117th Congress, when Democrats controlled the presidency, Senate, and House -- a political environment that may have incentivized progressives to register more protest votes against their party's legislative priorities. Similar patterns emerge for certain conservative Republicans, though with less visual prominence in the two-dimensional representation.

The changes in legislator positioning, while significant, appear less dramatic than those observed in our one-dimensional model. This moderation occurs because the additional dimension provides alternative ways to accommodate non-conforming votes beyond simply shifting legislators along the primary liberal-conservative spectrum. When a second dimension is available, the model can partially absorb protest voting behavior through movement in this additional dimension rather than through more extreme positioning on the primary ideological dimension.

By definition, a protest vote represents a strategic decision to vote against one's copartisans, manifesting as movement in the opposite direction along the primary ideological dimension. Multi-dimensional models, while mathematically more flexible, allow non-conforming votes to be mapped in various directions within the expanded space. This flexibility means that when protest votes can be positioned anywhere outside the main partisan bloc, the model may not always preserve the oppositional nature of protest voting along the primary dimension.

We present both one- and two-dimensional models as complementary approaches, each with distinct advantages. The one-dimensional \texttt{emRIRT\_L0} offers conceptual clarity and directional precision in capturing protest voting behavior, while the two-dimensional extension provides a more flexible and nuanced representation that accommodates greater complexity in voting patterns. For researchers primarily interested in understanding protest voting as a strategic behavior, the one-dimensional model offers a focused analytical framework. Meanwhile, for those investigating whether protest votes might reflect genuinely distinct secondary dimensions of preference, the two-dimensional approach provides valuable additional insights.

Our method demonstrates consistency across dimensionality in identifying protest voters, confirming the robustness of our approach. While the one-dimensional framework captures the directional nature of protest voting with particular clarity, the two-dimensional results enhance our understanding by revealing how protest voting behavior might interact with other dimensions of legislative preference. Both approaches confirm our method's ability to detect protest voting patterns, with each offering distinct perspectives that can enrich analyses of strategic legislative behavior.

\section{Conclusion}\label{sec_8}
Our study introduces a novel approach to ideal point estimation that explicitly accounts for protest voting behavior in legislative settings. By incorporating an $\ell_0$-regularized shift parameter, our method effectively identifies and adjusts for strategic votes that deviate from legislators' underlying ideological positions. The application to the 116th and 117th U.S. House of Representatives demonstrates that conventional methods can significantly underestimate ideological extremity when legislators engage in protest voting, particularly among members known for strategic voting behavior such as the Squad and certain Freedom Caucus members.

The method's effectiveness is validated through both simulation studies and empirical tests using documented protest votes. Our simulation results show that while traditional approaches and $\ell_1$-regularized alternatives suffer from increasing estimation error as protest voting increases, our $\ell_0$-regularized approach maintains consistent accuracy. The empirical validation using known protest votes reveals that our method conservatively identifies protest votes, distinguishing between genuine strategic behavior and general noise in voting patterns.

Beyond its methodological contribution, our approach provides important insights into legislative behavior. The systematic detection of protest votes across the ideological spectrum – from progressive Democrats to conservative Republicans – suggests that strategic voting is a broader phenomenon than previously documented. This finding has implications for our understanding of legislative dynamics and party discipline, particularly in an era of increasing polarization.

The framework we develop here opens several promising directions for future research. First, our method could be extended to analyze temporal patterns in protest voting, potentially revealing how strategic behavior evolves over time and across different congressional sessions. Second, the identification of protest votes could facilitate more detailed studies of when and why legislators choose to engage in strategic voting. Finally, our approach to robust ideal point estimation might be adapted to other contexts where strategic behavior can mask true preferences, such as judicial decision-making or international relations.

In conclusion, by developing a method that accurately captures both sincere and strategic voting behavior, we contribute to a more nuanced understanding of legislative decision-making. Our findings suggest that accounting for protest votes is crucial for accurate ideological scaling.

\clearpage
\bibliographystyle{chicago}
\bibliography{SADM}

\appendix

\section{Proofs of the Theorems in the Main Text}

\subsection{Proof of Theorem 1}\label{Appen_A.1}
Let $\boldsymbol{\vartheta}' = \{\theta_i\} \cup \{\boldsymbol{\tilde{\beta}}_j\} \cup \{\tau_{ij}\} \cup \{\kappa_{ij}\}$ 
and $\boldsymbol{\vartheta} = \{\theta_i\} \cup \{\boldsymbol{\tilde{\beta}}_j\} \cup \{\gamma_{ij}\}$. Consider
\vspace{-0.7mm}
\begin{align*}
Q_1(\boldsymbol{\vartheta}')
&= \sum_{i=1}^I \sum_{j=1}^J 
\left[
\zeta_{ij} 
- \log\!\left( \frac{1-\pi}{\pi} \right) \tau_{ij}
- \frac{1}{2} \left\{ y_{ij}^* - \left( \boldsymbol{\tilde{\beta}}_j^\top \boldsymbol{\tilde{\theta}}_i 
+ \tau_{ij}\kappa_{ij} \right) \right\}^2
\right], 
\\
Q_2(\boldsymbol{\vartheta})
&= \sum_{i=1}^I \sum_{j=1}^J 
\left[
\zeta_{ij} 
- \frac{\lambda^2}{2} \mathbb{I}(\gamma_{ij}\neq 0)
- \frac{1}{2} \left\{ y_{ij}^* - \left( 
\boldsymbol{\tilde{\beta}}_j^\top \boldsymbol{\tilde{\theta}}_i 
+ \gamma_{ij} \right) \right\}^2
\right].
\end{align*}

\vspace{-0.7mm}
First, suppose that $\max_{\boldsymbol{\vartheta}'} Q_1(\boldsymbol{\vartheta}') > \max_{\boldsymbol{\vartheta}} Q_2(\boldsymbol{\vartheta})$, and let $\boldsymbol{\vartheta}'^{\,*} = \{\theta_i^*\} \cup \{\boldsymbol{\tilde{\beta}}_j^*\} \cup \{\tau_{ij}^*\} \cup \{\kappa_{ij}^*\}$ be a maximizer of $Q_1$.  
Define $\gamma_{ij}^* = \tau_{ij}^* \kappa_{ij}^*$ and consider the induced parameter set $\boldsymbol{\vartheta}^* = \{\theta_i^*\} \cup \{\boldsymbol{\tilde{\beta}}_j^*\} \cup \{\gamma_{ij}^*\}$.  
By construction, $Q_1(\boldsymbol{\vartheta}'^{\,*})$ and $Q_2(\boldsymbol{\vartheta}^*)$ differ only in the second term of the summand.  
Since $\mathbb{I}(\gamma_{ij}^* \neq 0) = \mathbb{I}(\tau_{ij}^*\kappa_{ij}^*\neq 0) \le \tau_{ij}^*$, it follows that $Q_1(\boldsymbol{\vartheta}'^{\,*}) \le Q_2(\boldsymbol{\vartheta}^*)$, which implies $\max_{\boldsymbol{\vartheta}'} Q_1(\boldsymbol{\vartheta}') \le \max_{\boldsymbol{\vartheta}} Q_2(\boldsymbol{\vartheta})$, a contradiction.

Next, suppose that $\max_{\boldsymbol{\vartheta}'} Q_1(\boldsymbol{\vartheta}') < \max_{\boldsymbol{\vartheta}} Q_2(\boldsymbol{\vartheta})$, and let $\boldsymbol{\vartheta}^* = \{\theta_i^*\} \cup \{\boldsymbol{\tilde{\beta}}_j^*\} \cup \{\gamma_{ij}^*\}$ be a maximizer of $Q_2$.  
Define $\tau_{ij}^* = \mathbb{I}(\gamma_{ij}^*\neq 0)$ and $\kappa_{ij}^* = \gamma_{ij}^*$, and consider the induced parameter set $\boldsymbol{\vartheta}'^{\,*} = \{\theta_i^*\} \cup \{\boldsymbol{\tilde{\beta}}_j^*\} \cup \{\tau_{ij}^*\} \cup \{\kappa_{ij}^*\}$.  
By construction, $Q_1(\boldsymbol{\vartheta}^{\prime *})$ and $Q_2(\boldsymbol{\vartheta}^*)$ differ only in the third term of the summand. Since $\tau_{ij}^* \kappa_{ij}^* = \mathbb{I}(\gamma_{ij}^*\neq 0)\gamma_{ij}^* = \gamma_{ij}^*$, we have $Q_1(\boldsymbol{\vartheta}'^{\,*}) = Q_2(\boldsymbol{\vartheta}^*)$, which implies $\max_{\boldsymbol{\vartheta}'} Q_1(\boldsymbol{\vartheta}') \ge \max_{\boldsymbol{\vartheta}} Q_2(\boldsymbol{\vartheta})$, again a contradiction.

Therefore, we conclude that $\max_{\boldsymbol{\vartheta}'} Q_1(\boldsymbol{\vartheta}') = \max_{\boldsymbol{\vartheta}} Q_2(\boldsymbol{\vartheta})$, and the desired result follows.

\subsection{Proof of Theorem 2}\label{Appen_A.2}
\begin{proof}
     We begin by proving that \eqref{eqn:eqn1} implies  
    \begin{align}\label{eqn:eqn3}  
    \alpha_{j,1} + \beta_{j,1} \theta_{i,1} + \gamma_{ij,1} = \alpha_{j,2} + \beta_{j,2} \theta_{i,2} + \gamma_{ij,2} \quad \text{for all } i \in \mathcal{I} \text{ and } j \in \mathcal{J}.  
    \end{align}  
    For fixed indices $ i^\prime $ and $ j^\prime $, first consider the case where $ y_{i^\prime j^\prime} = 0 $. By \eqref{eqn:eqn1}, we obtain  
    \begin{align}\label{eqn:eqn4}  
    \begin{split}  
    &\left\{1-\Phi(\alpha_{j^\prime, 1} + \beta_{j^\prime, 1} \theta_{i^\prime, 1} + \gamma_{i^\prime j^\prime, 1})\right\} \times \prod_{i \neq i^\prime, j \neq j^\prime} p(y_{ij} \mid \alpha_{j,1}, \beta_{j,1}, \theta_{i,1}, \gamma_{ij, 1}) \\  
    = &\left\{1-\Phi(\alpha_{j^\prime, 2} + \beta_{j^\prime, 2} \theta_{i^\prime, 2} + \gamma_{i^\prime j^\prime, 2})\right\} \times \prod_{i \neq i^\prime, j \neq j^\prime} p(y_{ij} \mid \alpha_{j,2}, \beta_{j,2}, \theta_{i,2}, \gamma_{ij,2}).  
    \end{split}  
    \end{align}  
    Similarly, if we consider the case where $ y_{i^\prime j^\prime} = 1 $, we have  
    \begin{align}\label{eqn:eqn5}  
    \begin{split}  
    &\Phi(\alpha_{j^\prime, 1} + \beta_{j^\prime, 1} \theta_{i^\prime, 1} + \gamma_{i^\prime j^\prime, 1}) \times \prod_{i \neq i^\prime, j \neq j^\prime} p(y_{ij} \mid \alpha_{j,1}, \beta_{j,1}, \theta_{i,1}, \gamma_{ij, 1}) \\  
    = &\Phi(\alpha_{j^\prime, 2} + \beta_{j^\prime, 2} \theta_{i^\prime, 2} + \gamma_{i^\prime j^\prime, 2}) \times \prod_{i \neq i^\prime, j \neq j^\prime} p(y_{ij} \mid \alpha_{j,2}, \beta_{j,2}, \theta_{i,2}, \gamma_{ij,2}). 
    \end{split}  
    \end{align}  
    From equations \eqref{eqn:eqn4} and \eqref{eqn:eqn5}, we derive the following equality:  
    \begin{align*}  
    \dfrac{\Phi(\alpha_{j^\prime, 1} + \beta_{j^\prime, 1} \theta_{i^\prime, 1} + \gamma_{i^\prime j^\prime, 1})}{1-\Phi(\alpha_{j^\prime, 1} + \beta_{j^\prime, 1} \theta_{i^\prime, 1} + \gamma_{i^\prime j^\prime, 1})} = \dfrac{\Phi(\alpha_{j^\prime, 2} + \beta_{j^\prime, 2} \theta_{i^\prime, 2} + \gamma_{i^\prime j^\prime, 2})}{1-\Phi(\alpha_{j^\prime, 2} + \beta_{j^\prime, 2} \theta_{i^\prime, 2} + \gamma_{i^\prime j^\prime, 2})},  
    \end{align*}  
    which implies that  
    $
    \alpha_{j^\prime, 1} + \beta_{j^\prime, 1} \theta_{i^\prime, 1} + \gamma_{i^\prime j^\prime, 1} = \alpha_{j^\prime, 2} + \beta_{j^\prime, 2} \theta_{i^\prime, 2} + \gamma_{i^\prime j^\prime, 2}. 
    $
    Since $ i^\prime $ and $ j^\prime $ were chosen arbitrarily, this establishes our first claim. 

    To show that \eqref{eqn:eqn3} implies \eqref{eqn:eqn2}, define
    \begin{align*}
        \mathcal{I}_0 = \left\{ i \in \mathcal{I} : \gamma_{ij, 1} = \gamma_{ij, 2} = 0, \quad \forall j \in \mathcal{J} \right\}, \quad 
        \mathcal{J}_0 = \left\{ j \in \mathcal{J} : \gamma_{ij, 1} = \gamma_{ij, 2} = 0, \quad \forall i \in \mathcal{I} \right\}.
    \end{align*}
    Assumption~1 ensures that $\mathcal{J}_0$ is nonempty (indeed, $|\mathcal{J}_0| \ge 2$). For any $j \in \mathcal{J}_0$, we have
    \begin{align*}
        \alpha_{j,1} + \beta_{j,1} \theta_{i,1} + \gamma_{ij, 1} = \alpha_{j,2} + \beta_{j,2} \theta_{i,2} + \gamma_{ij, 2} 
        ~\Longleftrightarrow~ \alpha_{j,1} + \beta_{j,1} \theta_{i,1} = \alpha_{j,2} + \beta_{j,2} \theta_{i,2} \quad \forall i\in \mathcal{I}.
    \end{align*}
    In conjunction with Assumption~2, this implies that one of the following two cases holds:
    \begin{align}\label{eqn:res1}
    \begin{cases}
        (A)~ (\alpha_{j,1}, \beta_{j,1}, \theta_{i,1}) = (\alpha_{j,2}, \beta_{j,2}, \theta_{i,2}), &\text{for all } i \in \mathcal{I} \text{ and } j \in \mathcal{J}_0. \\ 
        (B)~  (\alpha_{j,1}, \beta_{j,1}, \theta_{i,1}) = (\alpha_{j,2}, -\beta_{j,2}, -\theta_{i,2}),  &\text{for all } i \in \mathcal{I} \text{ and }  j \in \mathcal{J}_0.
    \end{cases}
    \end{align}
    
    Now, we turn our attention to the case where $j \in \mathcal{J}_1 = \mathcal{J}\backslash \mathcal{J}_0 $. Consider a fixed $ j \in \mathcal{J}_1 $. By Assumption 1 and the definition of $\mathcal{I}_0$, there exist at least two elements $ i \in \mathcal{I}_0 $ satisfying
    \begin{align}\label{eqn:res2}
        \alpha_{j,1} + \beta_{j,1} \theta_{i,1} + \gamma_{ij, 1} = \alpha_{j,2} + \beta_{j,2} \theta_{i,2} + \gamma_{ij, 2}
        ~\Longleftrightarrow~ \alpha_{j,1} + \beta_{j,1} \theta_{i,1} = \alpha_{j,2} + \beta_{j,2} \theta_{i,2}.
    \end{align}
    To be concrete, let $ \mathcal{I}_0(j) $ denote the index set of values $ i $ that satisfy the above equivalence for a given $ j $. By rearranging the right-hand side of \eqref{eqn:res2} and using \eqref{eqn:res1}, it can be shown that one of the following linear systems holds:
    \begin{align}\label{eqn:res3}
    \begin{cases}
       (A)~ (\alpha_{j,1} - \alpha_{j,2}) + (\beta_{j,1} - \beta_{j,2}) \theta_{i,1} = 0, & \forall i \in \mathcal{I}_0(j).\\
       (B)~ (\alpha_{j,1} - \alpha_{j,2}) + (\beta_{j,1} + \beta_{j,2}) \theta_{i,1} = 0, & \forall i \in \mathcal{I}_0(j).\\
    \end{cases}
    \end{align}
    Since for all $ j \in \mathcal{J}_1 $, $ |\mathcal{I}_0(j)| \geq 2 $, one of the following holds:
    \begin{align}\label{eqn:res4}
        \begin{cases}
            (A)~ (\alpha_{j,1}, \beta_{j,1}) = (\alpha_{j,2}, \beta_{j,2}), & \forall j \in \mathcal{J}_1.\\
            (B)~ (\alpha_{j,1}, \beta_{j,1}) = (\alpha_{j,2}, -\beta_{j,2}) & \forall j \in \mathcal{J}_1.
        \end{cases}
    \end{align}
    Combining the results in \eqref{eqn:res1} and \eqref{eqn:res4}, we can conclude that \eqref{eqn:eqn3} implies \eqref{eqn:eqn2}.
\end{proof}

\subsection{Proof of Theorem 3}\label{Appen_A.3}
\begin{proof}
Following the same argument as in the proof of Theorem~\ref{theorem2}, 
equation~\eqref{eqn:eqn1_MD} implies
\begin{align}\label{eqn:eqn3_MD}
    \alpha_{j,1} + \boldsymbol{\beta}_{j,1}^\top \boldsymbol{\theta}_{i,1} + \gamma_{ij,1}
    =
    \alpha_{j,2} + \boldsymbol{\beta}_{j,2}^\top \boldsymbol{\theta}_{i,2} + \gamma_{ij,2},
    \qquad 
    \forall i\in\mathcal{I},\; \forall j\in\mathcal{J}.
\end{align}
Therefore, it suffices to show that \eqref{eqn:eqn3_MD} implies \eqref{eqn:eqn2_MD}.

Define $\mathcal{I}_0 = \{ i\in\mathcal{I} : \gamma_{ij,1}=\gamma_{ij,2}=0,\ \forall j\in\mathcal{J}\}$, $\mathcal{J}_0 = \{ j\in\mathcal{J} : \gamma_{ij,1}=\gamma_{ij,2}=0,\ \forall i\in\mathcal{I}\}$, and set $\mathcal{J}_1=\mathcal{J}\setminus\mathcal{J}_0$.
Assumption~1 ensures $|\mathcal{J}_0|\ge K+1$.
For every $j\in\mathcal{J}_0$, \eqref{eqn:eqn3_MD} simplifies to
\begin{equation*}
    \alpha_{j,1} + \boldsymbol{\beta}_{j,1}^\top\boldsymbol{\theta}_{i,1}
    =
    \alpha_{j,2} + \boldsymbol{\beta}_{j,2}^\top\boldsymbol{\theta}_{i,2},
    \qquad \forall i\in\mathcal{I}.
\end{equation*}
Let $\Theta_l$ be the $I\times K$ matrix whose $i$-th row is $\boldsymbol{\theta}_{i,l}^\top$.
Since $\Theta_l^\top \boldsymbol{1}_{I\times 1}=\boldsymbol{0}_{K\times 1}$ by Assumption~2,
summing over $i$ yields
\begin{equation*}
    \alpha_{j,1}=\alpha_{j,2}
    \quad\text{and}\quad
    \boldsymbol{\beta}_{j,1}^\top\boldsymbol{\theta}_{i,1}
    =
    \boldsymbol{\beta}_{j,2}^\top\boldsymbol{\theta}_{i,2},
    \qquad \forall i\in\mathcal{I},\ \forall j\in\mathcal{J}_0.
\end{equation*}
Because $|\mathcal{J}_0|\ge K+1$, one may select $K$ indices from $\mathcal{J}_0$ to construct full-rank
$K\times K$ matrices $B_l$ whose rows are $\boldsymbol{\beta}_{j,l}$, $l\in\{1,2\}$.
The preceding identity implies
\begin{equation*}
    B_1\boldsymbol{\theta}_{i,1} = B_2\boldsymbol{\theta}_{i,2},
    \qquad \forall i\in\mathcal{I}.
\end{equation*}
Let $O=B_1^{-1}B_2$. Then $\boldsymbol{\theta}_{i,1}=O\boldsymbol{\theta}_{i,2}$ for all $i\in\mathcal{I}$, and hence $\Theta_1=\Theta_2O^\top$.
Assumption~2 also gives $I^{-1}\Theta_l^\top\Theta_l=I_{K\times K}$ for $l\in\{1,2\}$.  
Therefore,
\begin{equation*}
    I_{K\times K}
    =
    I^{-1}\Theta_1^\top\Theta_1
    =
    O(I^{-1}\Theta_2^\top\Theta_2)O^\top
    =
    OO^\top,
\end{equation*}
showing that $O$ is orthogonal. Thus, we have
\begin{equation*}
    \alpha_{j,1}=\alpha_{j,2},
    \qquad
    \boldsymbol{\beta}_{j,1}=O\boldsymbol{\beta}_{j,2},
    \qquad
    \boldsymbol{\theta}_{i,1}=O\boldsymbol{\theta}_{i,2},
    \quad \forall i\in\mathcal{I},\; \forall j \in \mathcal{J}_0.
\end{equation*}

Now consider $j\in\mathcal{J}_1$ and define $\mathcal{I}_0(j) = \bigl\{i\in\mathcal{I} :\alpha_{j,1} + \boldsymbol{\beta}_{j,1}^\top\boldsymbol{\theta}_{i,1} = \alpha_{j,2} + \boldsymbol{\beta}_{j,2}^\top\boldsymbol{\theta}_{i,2} \bigr\}$.
Substituting $\boldsymbol{\theta}_{i,1}=O\boldsymbol{\theta}_{i,2}$ gives
\begin{equation*}
    (\alpha_{j,1}-\alpha_{j,2})
    +
    (O^\top\boldsymbol{\beta}_{j,1}-\boldsymbol{\beta}_{j,2})^\top\boldsymbol{\theta}_{i,2}
    =0,
    \qquad \forall i\in\mathcal{I}_0(j).
\end{equation*}
Assumption~1 ensures $|\mathcal{I}_0(j)|\ge K+1$, and hence the only solution is
\begin{equation*}
    \alpha_{j,1}=\alpha_{j,2}
    \quad\text{and}\quad
    O^\top\boldsymbol{\beta}_{j,1}=\boldsymbol{\beta}_{j,2},
\end{equation*}
Thus, for all $j\in\mathcal{J}$ and $i\in\mathcal{I}$,
\begin{equation*}
    (\alpha_{j,1},\,\boldsymbol{\beta}_{j,1}^\top,\,\boldsymbol{\theta}_{i,1}^\top)
    =
    (\alpha_{j,2},\,(O\boldsymbol{\beta}_{j,2})^\top,\,(O\boldsymbol{\theta}_{i,2})^\top),
\end{equation*}
which establishes \eqref{eqn:eqn2_MD}.
\end{proof}

\section{Implementation Details of Algorithm \ref{alg:EM}}\label{Appen_B}
In this section, we outline the implementation details of the EM-algorithm described in Algorithm \ref{alg:EM}. To execute the EM algorithm, it is necessary to specify the hyperparameters $\lambda$, $\mu_{\theta}$, $\sigma_\theta^2$, $\boldsymbol{\mu}_{\boldsymbol{\tilde{\beta}}}$, and $\Sigma_{\boldsymbol{\tilde{\beta}}}$, along with the initial estimate $\boldsymbol{\vartheta}^{(0)}$.

The hyperparameter $\lambda$ controls the sparsity level of the matrix $\Gamma$. Larger values of $\lambda$ lead to a sparser $\Gamma$. In particular, setting $\lambda = \infty$ results in $\Gamma = 0$, thereby reducing our model to that proposed by \cite{Imai2016}. We recommend using $\lambda = 3$ as the default value, based on empirical observation. 

\begin{figure}[H]
    \centering
    \includegraphics[width=0.9\linewidth]{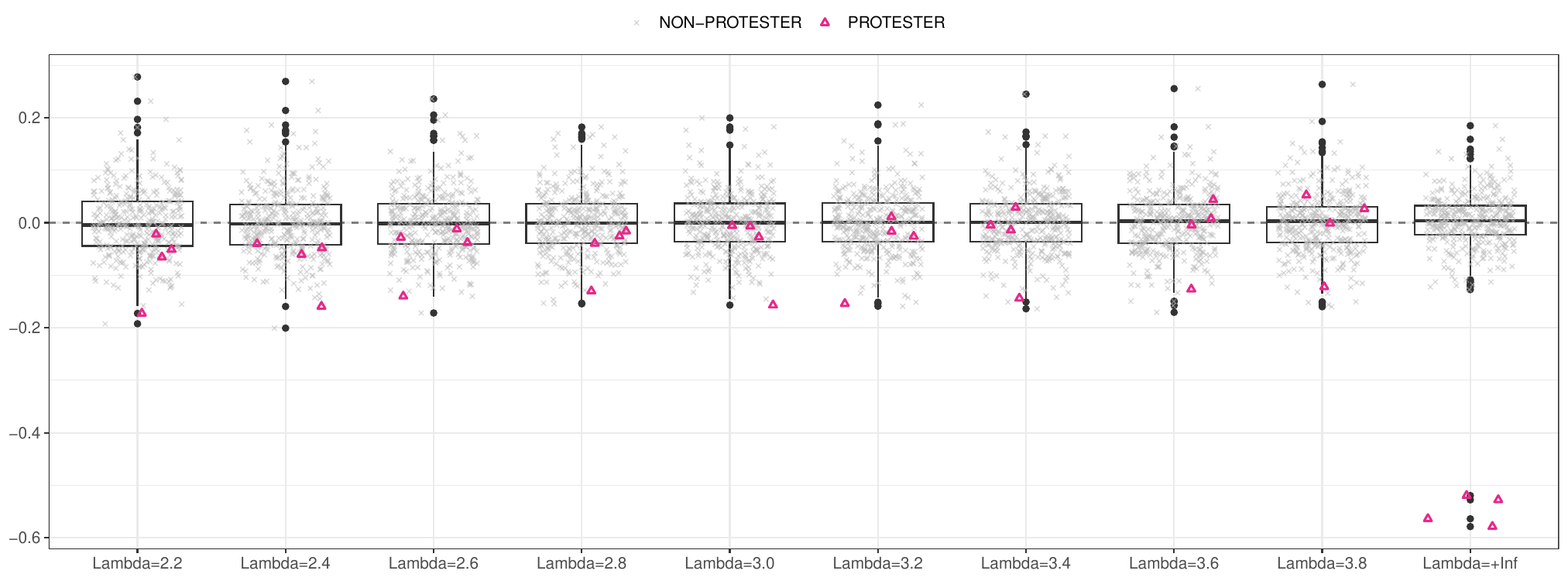}
    \caption{Boxplot of estimation errors in ideal points across varying values of $\lambda$. The x-axis represents the value of $\lambda$ used to estimate the ideal points, while the y-axis shows the estimation error (true ideal point minus estimated ideal point). Grey crosses denote the estimation errors for non-protesters, and pink triangles denote those for the four protesters. The points are horizontally jittered to reduce overplotting.}
    \label{fig:choice_lambda}
\end{figure}

To examine how sensitive our method is to the choice of $\lambda$, we conduct a simulation study with 395 legislators, 1{,}000 bills, four protesters each casting 40 protest votes, and $\lambda$ varying from 2.2 to 3.8 in increments of 0.2. Figure~\ref{fig:choice_lambda} shows that performance is remarkably stable across this range. For all values $\lambda \in \{2.2, 2.4, \ldots, 3.8\}$, estimation errors for protesters remain within $[-0.2, 0.2]$—substantially better than the benchmark case where $\lambda = +\infty$ (equivalent to \texttt{emIRT}), which produces errors below $-0.5$ for protesters. This stability suggests that practitioners can use the recommended default value $\lambda = 3$ with confidence, without extensive tuning.

For the hyperparameters $\mu_{\theta}$, $\sigma_\theta^2$, $\boldsymbol{\mu}_{\boldsymbol{\tilde{\beta}}}$ and $\Sigma_{\boldsymbol{\tilde{\beta}}}$, we follow the recommendations in \cite{Clinton2004}. Specifically, we set $\mu_{\theta} = 0$, $\sigma_\theta^2 = 1$, $\boldsymbol{\mu}_{\boldsymbol{\tilde{\beta}}} = (0, 0)^\top$, and $\Sigma_{\boldsymbol{\tilde{\beta}}} = 25 I_{2\times2}$, where $I_{K \times K}$ denotes the $K\times K$ identity matrix.

The choice of initial estimates $\boldsymbol{\vartheta}^{(0)}$ is important for ensuring the robustness of the final parameter estimates. We recommend performing a preliminary run to obtain an initial estimate of $\boldsymbol{\vartheta}$ using the following choice of hyperparameters and initial values $\boldsymbol{\vartheta}^{(0)}$: All hyperparameters are set to their default values, except that $\lambda$ is set to 2 instead of 3. The initial values $\boldsymbol{\vartheta}^{(0)}$ for the preliminary run are generated as follows: $\theta_i^{(0)}$, $\alpha_j^{(0)}$, and $\beta_j^{(0)}$ are drawn from a standard normal distribution, and $\gamma_{ij}^{(0)}$ is initialized to 0 for all $i = 1, 2, \ldots, I$ and $j = 1, 2, \ldots, J$. 

While these choices of hyperparameters and initialization may not be optimal in all cases, they have demonstrated consistent performance across all simulation settings. 

\section{Preprocessing of the Roll Call Data}\label{Appen_C}
To ensure data quality and consistency, we followed standard preprocessing steps commonly used in roll call analysis. The procedure involved a sequential screening process. First, we filtered out bills where the minority position received 1\% or fewer of the total votes, as such near-unanimous decisions provide little information for ideal point estimation. Next, we excluded legislators who participated in fewer than 10\% of the remaining roll calls, since sparse voting records can result in unstable estimates. Finally, we removed unanimously passed bills, which do not help differentiate legislators' ideological positions.

For the 116th House of Representatives, we additionally excluded Representative Justin Amash prior to applying the preprocessing steps, due to his mid-term party switch from Republican to Independent. This exclusion is consistent with standard practices in the literature when dealing with party-switching legislators.

\FloatBarrier
\section{Validity Assessment of the One-Dimensional \texttt{emRIRT\_L0} Method : 116th House of Representatives}\label{Appen_D}

To validate our method's effectiveness in identifying protest votes, we conduct an empirical test using documented cases of protest voting behavior. \cite{Lewis2022} presented a list of 13 roll call votes in the 116th Congress where Squad members cast likely protest votes, providing us with ``ground truth'' cases to evaluate our method's detection capabilities. This test is particularly valuable as it allows us to assess whether \texttt{emRIRT\_L0} can identify already-documented instances of protest voting, serving as an important check on the method's real-world performance. Of these 13 roll calls (RC 72, 86, 346, 413, 450, 630, 651, 805, 851, 877, 905, 937, and 949), \texttt{emRIRT\_L0} identified nine as protest votes ($\gamma_{ij} \neq 0$ for RC 72, 86, 346, 413, 630, 651, 805, 877, and 905) and failed to identify four ($\gamma_{ij} = 0$ for RC 450, 851, 937, and 949).

\begin{figure}[!htb]
    \centering
    \includegraphics[width=0.9\linewidth]{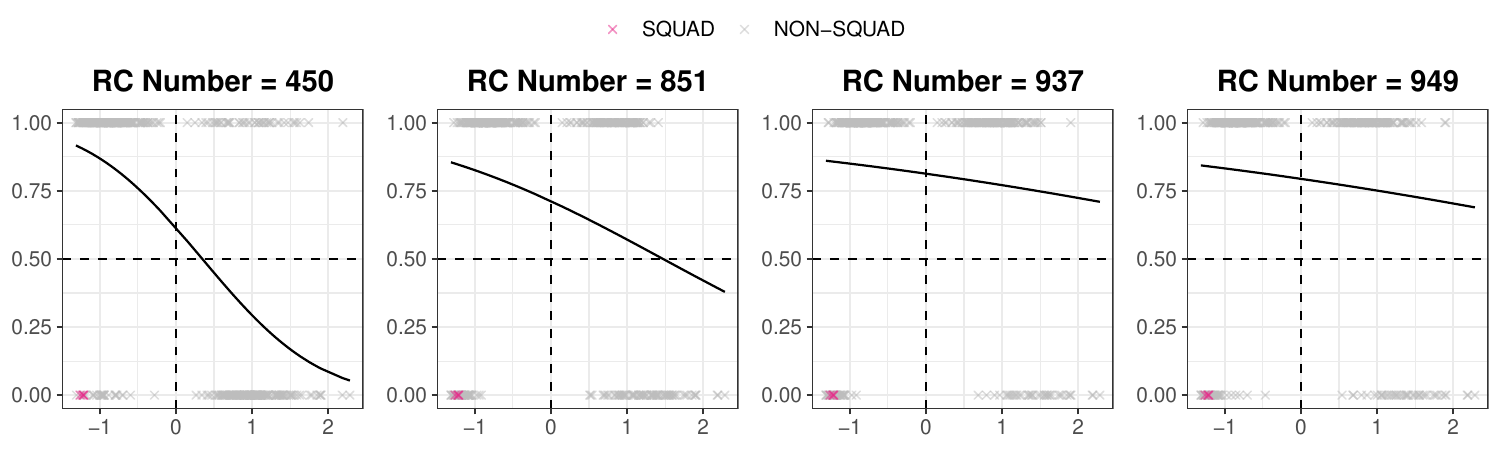}
    \caption{Item response curves for bills where Squad members' protest votes were not detected. The $x$-axis shows estimated ideal points and the $y$-axis shows the probability of voting yea. Black solid lines represent item response curves, with grey crosses at $y = 0$ or $y = 1$ showing individual voting results. Pink crosses indicate Squad members' votes, and blue diamonds show votes identified as protests by \texttt{emRIRT\_L0} (i.e. $\gamma_{ij} \neq 0$).}
    \label{fig:undet}
\end{figure}

\begin{figure}[!htb]
    \centering
    \includegraphics[width=0.7\linewidth]{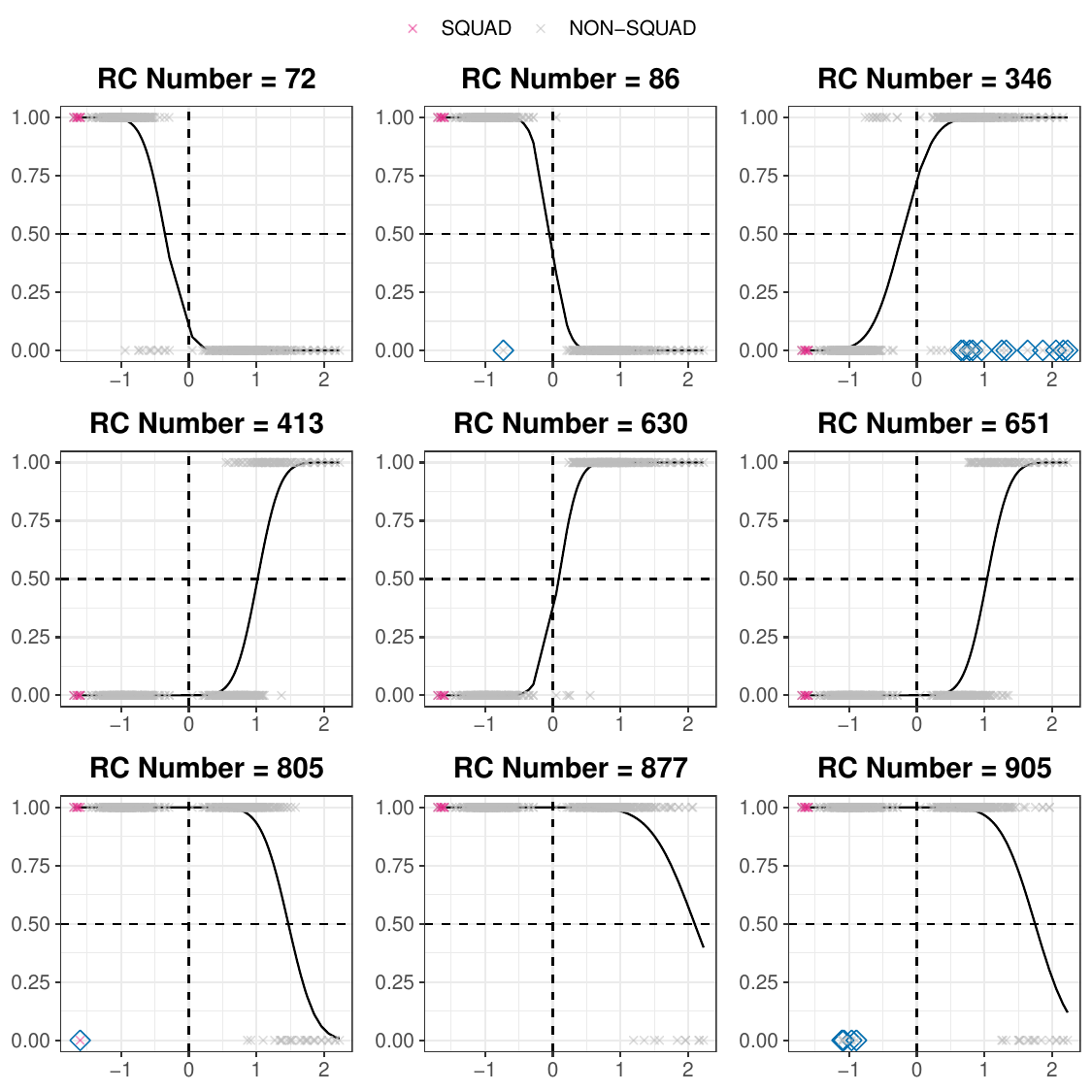}
    \caption{Item response curves for bills where Squad members' protest votes were detected. The $x$-axis shows estimated ideal points and the $y$-axis shows the probability of voting yea. Black solid lines represent item response curves, with grey crosses at $y = 0$ or $y = 1$ showing individual voting results. Pink crosses indicate Squad members' votes, and blue diamonds show votes identified as protests by \texttt{emRIRT\_L0} (i.e. $\gamma_{ij} \neq 0$).}
    \label{fig:det}
\end{figure}

At first glance, the detection of only nine out of thirteen known protest votes might suggest limited performance of our method. However, a careful examination of the item response curves reveals that \texttt{emRIRT\_L0} appropriately distinguishes between different types of voting deviations. To understand these results in detail, we analyze the estimated item response curves from \texttt{emRIRT\_L0}. Figures \ref{fig:undet} and \ref{fig:det} show the curves (black solid lines) for the unidentified and identified protest votes, respectively.

In all thirteen bills, the Squad members' votes (shown as pink crosses) deviate from the expected outcomes predicted by the item response curves - their observed nay votes (y = 0) occur in regions where the model predicts high probability of yea votes. This pattern illustrates how protest votes fundamentally challenge conventional ideal point estimation methods by producing behavior that lies outside the expected ideological voting patterns.

The key to understanding why our method detects some cases but not others lies in the $\gamma_{ij}$ parameter, which acts as a regularizing constraint similar to a spike-and-slab prior in Bayesian variable selection. For the four undetected roll calls (Figure \ref{fig:undet}), although the Squad members' votes deviate from the curve's predictions, similar deviations are observed among other legislators, as evidenced by the scattered pattern of both yea and nay votes across the ideological spectrum. In all cases, the $\ell_0$ regularization sets $\gamma_{ij} = 0$ for Squad members because their voting behavior, while unexpected, is not sufficiently distinct from the overall noise in the voting pattern to warrant classification as protest votes.

In contrast, for the nine detected roll calls (Figure \ref{fig:det}), the Squad members' votes represent stark outliers from an otherwise monotonic relationship between ideology and voting behavior. For example, RC 86 involved appropriations for the Department of Homeland Security, RC 630 concerned appropriations for the Departments of Commerce and Justice, and RC 651 addressed a resolution regarding U.S. efforts in the Israeli-Palestinian conflict. In these cases, the Squad's opposition stood out as distinct from the broader voting patterns of even their progressive colleagues. The $\ell_0$ regularization identifies these votes as requiring protest vote parameters ($\gamma_{ij} \neq 0$) because they cannot be explained by the general pattern of ideological voting, even accounting for noise in the response curve.

This pattern is particularly evident in RC 905, which concerned the Expanding Access to Sustainable Energy Act, and RC 877, which dealt with Department of Defense appropriations. In both cases, the bills received broad bipartisan support (with respective probabilities of 99.6\% and 100\% for supporting votes from legislators with similar ideological positions), making the Squad's opposition statistically distinctive. This distinction demonstrates that our method is conservative in its classification, identifying protest votes only when they represent statistically significant departures from the underlying voting pattern that cannot be absorbed into the general variability of the item response curve.

\FloatBarrier
\section{Additional Application : 118th House of Representatives}\label{Appen_E}
\begin{figure}[!hbt]
    \centering

    \begin{overpic}[width=0.4\linewidth]{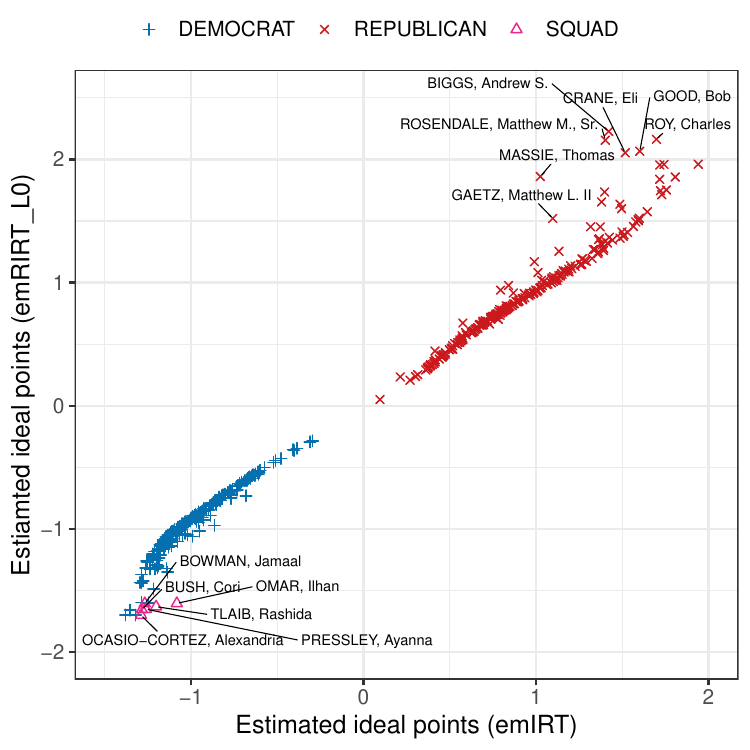}
        \put(2,95){\textbf{(A)}}
    \end{overpic}
    
    \vspace{0.8cm}

    \begin{overpic}[width=0.8\linewidth]{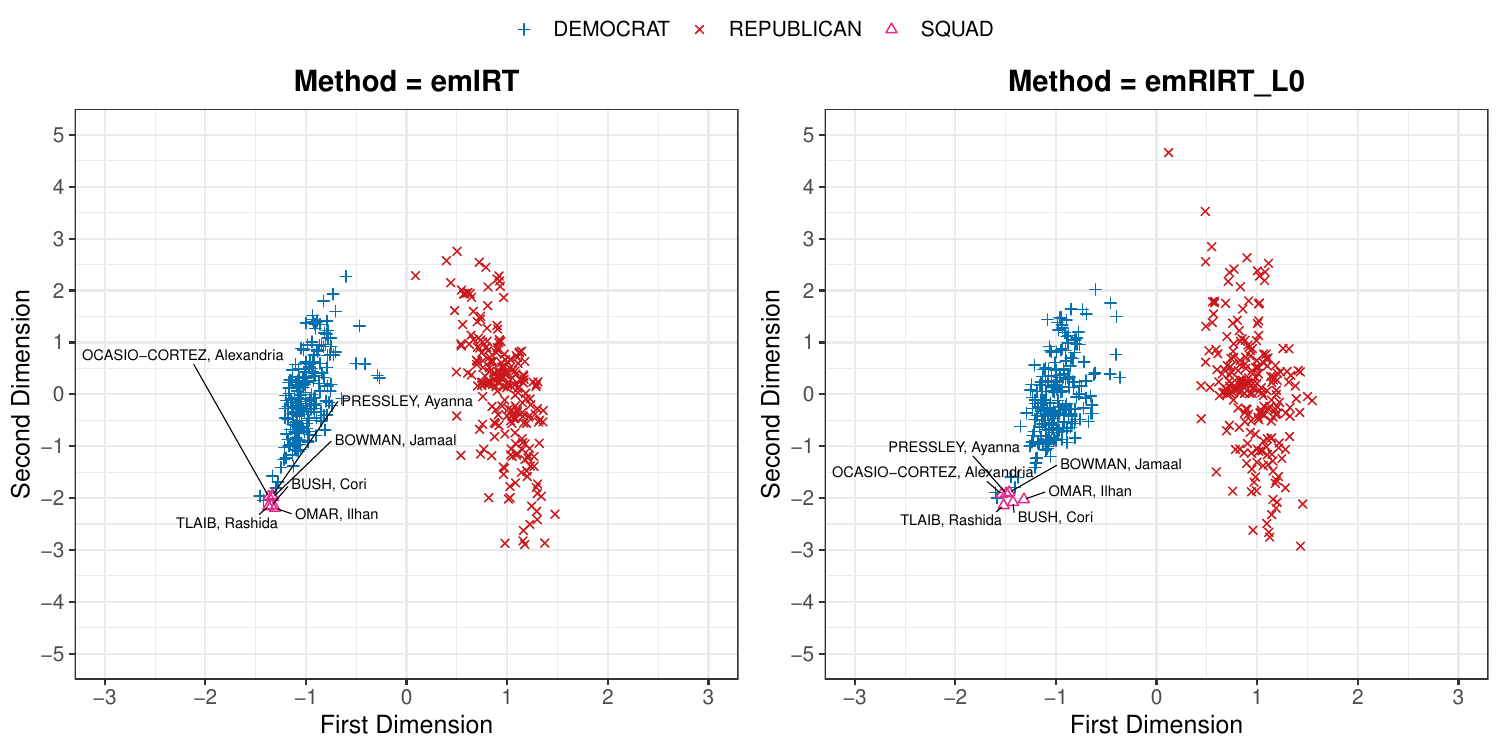}
        \put(2,47){\textbf{(B)}}
    \end{overpic}

    \caption{Comparison of ideal point estimates for the 118th Congress using two methodological approaches. Panel A presents results from a one-dimensional model, where the $x$-axis represents ideal point estimates via \texttt{emIRT} and the $y$-axis shows estimates from \texttt{emRIRT\_L0}. Panel B presents results from a two-dimensional model, where the left plot shows estimates from \texttt{emIRT} and the right plot from \texttt{emRIRT\_L0}. In all plots, Democratic members are marked with blue plus signs, Republican members with red crosses, and Squad members are highlighted with pink triangles.}
    \label{fig:H118}
\end{figure}

In this section, we apply both one-dimensional and two-dimensional versions of $\texttt{emRIRT\_L0}$ and $\texttt{emIRT}$ to the roll call data from the 118th U.S. House of Representatives. Prior to the analysis, we follow the standard preprocessing procedure described in Appendix \ref{Appen_B}. The results are presented in Figure \ref{fig:H118}.

\FloatBarrier
\section{Implications for Pivotal Quantities}\label{Appen_F}
\begin{figure}[!htb]
\centering
\includegraphics[width=0.9\linewidth]{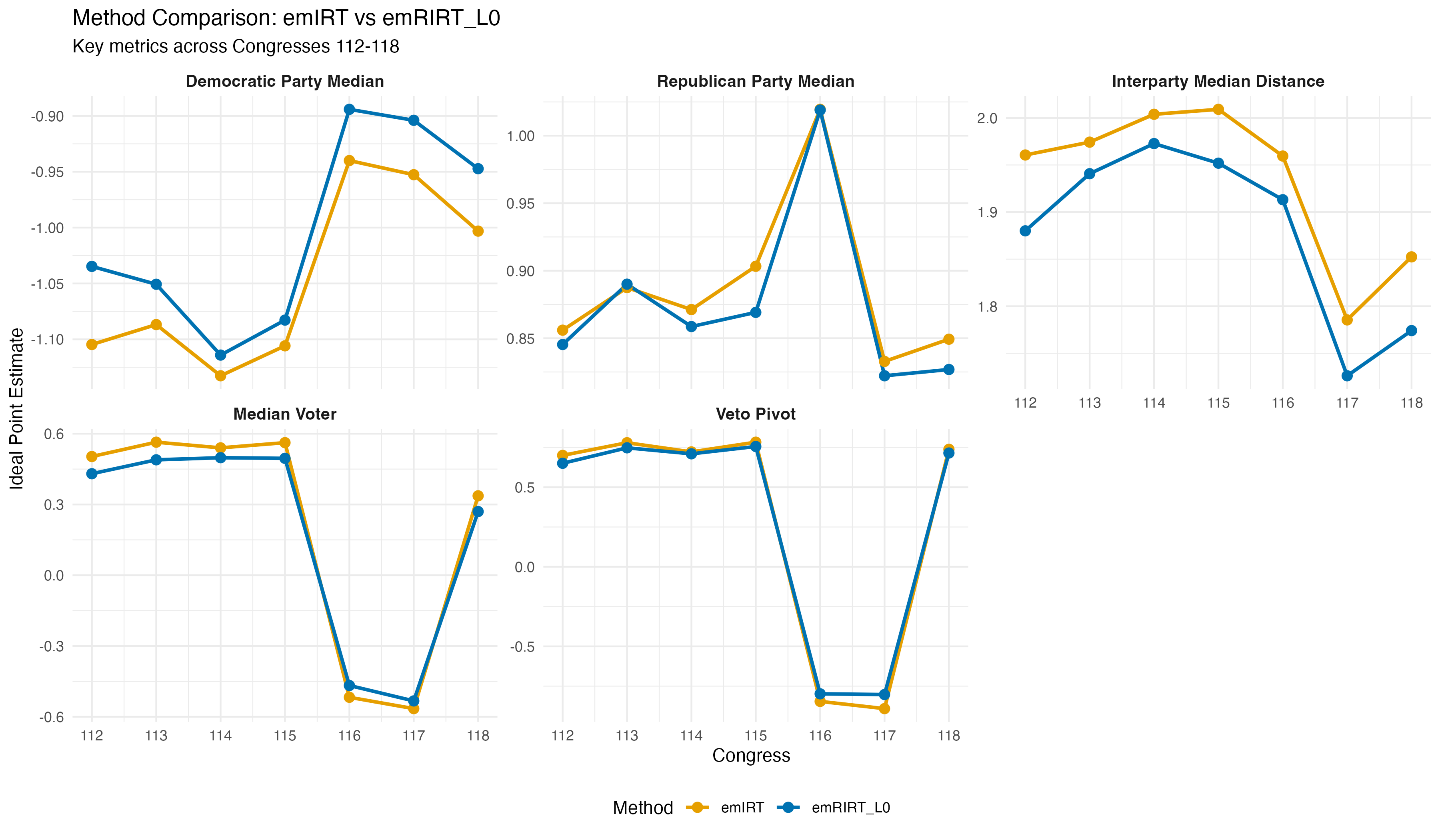}
\caption{Estimates of pivotal quantities of the U.S. Congress from the 112th to 118th using \texttt{emIRT} (yellow) and \texttt{emRIRT\_L0} (blue).}
\label{fig:prac}
\end{figure}

To assess the broader implications of accounting for protest votes, we examine how \texttt{emRIRT\_L0} affects estimates of pivotal quantities commonly used in legislative studies \citep{krehbiel1998pivotal}. Figure \ref{fig:prac} compares five key metrics across the 112th to 118th Congresses using \texttt{emIRT} (yellow) and \texttt{emRIRT\_L0} (blue).
First, the upper-left and upper-middle panels display party median estimates over time. While our method produces substantial repositioning of individual legislators who engage in protest voting (as documented in Section \ref{sec_61}), the effects on party medians are more nuanced. Democratic party medians show consistent but modest differences between methods, while Republican party medians remain largely similar across most congresses, with notable exceptions in the 115th and 118th.

Second, the upper-right panel presents interparty median distance, a common measure of partisan polarization. The two methods yield somewhat different estimates of polarization, with differences ($\frac{|\texttt{emIRT} - \texttt{emRIRT\_L0}|}{\texttt{emIRT}} \times 100$) ranging from 1.6\% to 4.2\% across congresses. This finding shows that how to account for protest voting influences substantive conclusions about polarization trends—a quantity of central interest in contemporary legislative studies.

Third, and importantly, the floor median (Median Voter) and veto pivot estimates shown in the bottom panels exhibit remarkable stability across methods. This pattern is theoretically expected and provides indirect validation of our approach. By definition, protest votes are strategic: legislators cast them precisely when doing so will not alter legislative outcomes. This strategic logic implies that protest votes should affect within-party positioning without shifting cross-party pivotal quantities. The stability of floor median and veto pivot estimates confirms that our method identifies votes consistent with strategic behavior—votes that affect individual positioning but not the decisive votes that determine bill passage.

These results highlight an important distinction between individual-level and aggregate-level effects. Accounting for protest votes substantially changes our understanding of where individual legislators—particularly ideological activists like Squad members and Freedom Caucus members—stand relative to their copartisans. However, because protest votes are strategically cast to avoid affecting outcomes, their correction appropriately preserves estimates of quantities that determine legislative success and failure.

\end{document}